\documentclass[aps,pre]{revtex4-2}  
\usepackage[margin=1in]{geometry}
\usepackage{amsmath,amssymb,amsfonts,mathrsfs}
\usepackage{graphicx}
\usepackage{xcolor}
\usepackage{enumitem}

\newcommand{\kl}[2]{\text{D}(#1 \lVert #2)}
\newcommand{\tkl}[2]{\widetilde{\text{D}}(#1 \lVert #2)}
\newcommand{\bp}{\mathbf{p}}
\newcommand{\bq}{\mathbf{q}}

\newcommand{\qhat}{\widehat{q}}

\newcommand{\bqhat}{\widehat{\mathbf{q}}}
\newcommand{\ssn}{^{(n)}}
\newcommand{\ssl}{^{(\ell)}}
\newcommand{\bbar}{\overline{\beta}}
\newcommand{\gbar}{\overline{\gamma}}
\newcommand{\coast}{\texttt{COAST}}
\newcommand{\astxc}{A^{\text{XC}}}
\newcommand{\astuc}{A^{\text{UC}}}
\newcommand{\xclent}{\texttt{XclEnt}}
\newcommand{\decent}{\texttt{DecEnt}}
\newcommand{\thrent}{\texttt{ThrEnt}}
\newcommand{\totent}{\texttt{TotEnt}}

\begin{document}

\title{Routes to rare events with optimally timed perturbations: a Tent Map is all you need}

\author{Justin Finkel}
\email{jfinkel@uchicago.edu}
\affiliation{Data Science Institute and the Department of Geophysical Sciences, University of Chicago}

\begin{abstract}
    Extreme weather events are difficult to understand for the same reason that they are dangerous: they happen rarely, catching victims unprepared when they do occur and scientists unable to assess risks confidently, given such limited precedent to learn from in the real world and high computational expense to simulate more examples. Rare event sampling (RES) algorithms seek to reduce this expense by forcing simulations more directly towards the extremes and then compensating for that forcing in statistical analysis. But the performance of RES hinges on several hyperparameter choices which are ad hoc in practice, and must be better understood if RES is to be broadly useful. This paper addresses one particular parameter, the \emph{advance split time} (AST), which prescribes when to perturb a simulation to split off the most informative possible ensemble of alternative extreme event scenarios. We prescribe the optimal AST as the time it takes for an initial perturbation to amplify into the size (inverse rarity) of the extreme event being targeted. For the Logistic and Tent maps, two archetypal examples of one-dimensional chaos, we rigorously derive and express the rule as a simple log-ratio between perturbation size and event rarity. The pair of examples also illuminates where the rule breaks down, and subsequently, we generalize the rule into a maximum-entropy criterion that solidifies recent heuristic and empirical results. Despite the idealized setting, our results deliver theoretical clarity that can anchor future developments of principled RES methods applicable to real-world, high-impact weather and climate extremes. 
\end{abstract}

\maketitle

\section{Introduction}
\label{sec:introduction}

The traditional boundaries in climate science between physical and statistical models, between ``pure'' physical and ``applied'' impacts-oriented research, and between weather and climate timescales, are being blurred. A topic of intense current interest is extreme weather events, not only their detailed physical dynamics but their long-term occurrence frequencies and how they are changing \citep{Coumou2012decade,Fischer2021increasing,Faranda2024statistical}. As forecasts extend farther to seasonal and decadal horizons, a probabilistic framing becomes crucial and these extremes become inevitable, even if improbable on any given day \citep{White2017potential,Vitart2017subseasonal,Bloomfield2021subseasonal}. 

But ``probability'' has several interpretations. Ensemble \emph{weather} forecasts give subjective probabilities of near-future states given two uncertain inputs: the present state with partial and imprecise measurements, and the dynamical model with structural and parametric errors. Both uncertainties are properties of technology, not of the physical system being predicted, so the resulting forecast distributions can't be judged for ``correctness'' but only self-consistency, for example with reliability diagrams \citep{Wilks2019statistical_ch9}. 
\emph{Climate} probabilities on the other hand, which we focus on in this work, are objective. ``Climatology'', or ``steady-state'', refers to the statistics of a dynamical system accumulated over time in the long-term limit where the ergodic hypothesis applies and initial conditions don't matter \citep{Ghil2020physics,Patel2023using}. Of course the Earth doesn't attain this limit, having a finite history and changing boundary conditions, but the same dichotomy between weather and climate probabilities applies to model systems. For methodological development purposes, this work will assume a given explicit dynamical system, with fixed parameters ergo a stationary climate, as ``truth'', and take on the challenge not of improving its match to the physical world but of estimating climate probabilities of this fixed model. 

In principle, one can estimate climatological probabilities by ``Direct Sampling'' (DS): running a model autoregressively for a long enough time until its statistics converge to satisfactory precision. Yet in practice, obtaining that ground truth becomes exceedingly expensive when the quantities of interest pertain to extreme events. Correctly calculating a probability of 1/(250 years) from data is hard precisely because it is small. We can expect to see one such event in a 250-year timeseries (roughly the length of the longest-running temperature records), but the probability of seeing no events at all is $(1-\frac1{250})^{250}\approx e^{-1}$, almost as likely as seeing the expected one single event. In other words, getting an estimate 100\% off from the truth is almost as likely as getting it correct. More importantly, actual simulations of this extreme scenario---so-called ``storylines'', which are critical ingredients for risk mitigation and for scientific understanding \citep{Shepherd2018storylines,Sillmann2021event}--accrue at a painfully slow rate when integrating a model in DS mode. 

Rare event sampling (RES) has gained recent traction in climate science as a collection of computational protocols to accelerate difficult, high-impact estimation problems such as prolonged hot or wet seasons \citep{Ragone2018computation,Wouters2023rare,Noyelle2025statistical}, shorter-duration heat waves and storms \citep{Finkel2026rare,BloinWibe2025estimating,Lancelin2026ai}, intense tropical cyclones \citep{Webber2019practical}, and regime transitions like Atlantic Meridional Overturning Circulation collapse \citep{JacquesDumas2024estimation}, to name a few. RES aims to capture both statistics and storylines at reduced cost relative to DS by interrupting the forward time evolution with small, strategic perturbations on the model state to short-cut the long waiting periods and fast-forward to the extremes, while keeping track of the bias introduced by kicks to compensate for it afterward in statistical estimation. As in forecasting, the perturbations are often ad-hoc, with wildly divergent choices appearing in recent literature, from roundoff error in temperature \citep{Gessner2021very} to uniform noise on log-surface pressure spherical harmonics \citep{Ragone2018computation} to red noise forcing on parameterized subgrid tendencies \citep{Finkel2026rare}. But unlike in the weather forecasting task, these RES schemes target an actual unique, objective ground-truth distribution: namely, the climatology produced by DS. This makes some perturbation schemes objectively better than others. 

Our goal in this paper is to establish a principle for optimal perturbations, in the sense of recovering a climatological ground truth distribution both accurately and efficiently. The scope is narrow, with a focus on one specific rare event algorithm known as \emph{Ensemble Boosting} (EB) introduced in \cite{Gessner2021very}, only perturbation \emph{timing} as the thing to optimize given a fixed perturbation \emph{structure}, and a particular one-dimensional chaotic system known as the Logistic Map. But our solution is exact, expressible as a simple balance between the perturbation's initial size, its time horizon for growth, and the size in state space of the small rare-event target it aims for (the rarer, the smaller). The result is not universal, but is transparent in its assumptions. By first exploiting and then abstaining from a conjugate transformation to the Tent Map, we provide evidence that a \emph{maximum-entropy} principle could be the right way to generalize the result, amending a conjecture that we recently made in \cite{Finkel2026boosting} while refuting some other recently-proposed heuristics. Since EB is a very simple algorithm to implement and the perturbation timing is easy to adjust within complex pre-existing model architectures, we suspect that our simplified but illuminating results can be broadly useful. 

This paper is structured as follows. Section \ref{sec:ebwhite} recapitulates the EB procedure and identifies the open questions regarding its validity and performance. Section \ref{sec:logisticmap} introduces the Logistic Map, transforms to the Tent Map to derive and explain the exact solution to optimal perturbation timing, and then returns to the Logistic Map to demonstrate the potential generalizability of the maximum-entropy principle. Section \ref{sec:discussion} discusses the validity of the assumptions made and how they might be generalized, and provides a roadmap for putting the results to practical use. Section \ref{sec:conclusion} concludes the paper with a summary. 

\section{Ensemble Boosting general procedure}
\label{sec:ebwhite}

EB is used to study the rare, extreme fluctuations of a dynamical system $x(t)$ that evolves in time according to some update rule $x(t+1)=F(x(t))$, and which over long time periods behaves like a random variable filling out a nontrivial probability distribution, the ``climatology'', as opposed to converging to a fixed point or limit cycle. We seek to estimate the complementary cumulative distribution functions (CCDF) of the distribution tail, taking the form $Q(\nu,\mu):=\mathbb{P}\{x>\nu|x>\mu\}$, where $\mu$ is some pre-determined threshold separating the distributions ``body'' of typical events from its ``tail'' of extreme events and $\nu>\mu$ is a higher level within the tail. In the real world, $\mu$ might be the height of a dam, and $\nu$ might represent thresholds for even greater damage such as the height of a bridge or floodwall. (In general, extremes would be defined by some observable $R(x)$ of a high-dimensional state vector $x$, but in our simple 1D system we just choose $R(x)=x$.) 

EB tries to squeeze as much information as possible from the occasional event that appears in simulation. In general, we assume that the baseline threshold $\mu$ is just frequent enough that a moderately long DS can estimate the baseline exceedance probability, $\mathbb{P}\{x>\mu\}$, with reasonable confidence, and boosting targets everything above it. Rather than keep simulating the system forward and wait for new favorable initial conditions, EB goes back in time, perturbs the initial conditions already found, and re-simulates a new version of the event. In detail, the procedure is

\begin{enumerate}
    \item From an initial timeseries $x(t)$, find an exceedance event $x^*=x(t^*)>\mu$ (if there is a cluster of exceedances, choose $x^*$ as the peak). 
    \item Choose an advance split time $A$, and extract the initial condition $x(t^*-A)$  from the simulation. 
    \item Apply a small perturbation $\delta x$ to get a modified initial conditoin, $x(t^*-A)+\delta x$.
    \item Run the dynamics forward again through the original event timing, generating the short ``descendant'' timeseries of length $A+1$,
        \begin{align}
            \begin{bmatrix}
                x(t^*-A)+\delta x & F\big(x(t^*-A)+\delta x\big) & \hdots &  F^A\big(x(t^*-A)+\delta x\big)
            \end{bmatrix},
        \end{align}
        branching off of the ``ancestor'' timeseries at time $t^*-A$. 
    \item Repeat steps 2-4 to generate an ensemble of $D$ different perturbations $\{\delta x_d\}_{d=1}^D$, resulting in an ensemble of $D$ perturbed descendant peaks $\{x_d^*\}_{d=1}^D$ (we assume the perturbations are small enough to preserve the timing of local maxima, which is usually valid for lead times of interest in discrete-time systems but needs to be relaxed for continuous-time systems, as in \cite{Finkel2026boosting}).
    \item Repeat step 5 (and by recursion, 2-4) on a collection of ancestor events $\{x(t_n^*)\}_{n=1}^N$ drawn from the initial timeseries, resulting in $N$ ``families''. Enforce a minimum lag time between consecutive clusters to ensure independence between families.
    \item From the two-tiered population $\{x_{d,n}^*\}$, estimate tail probabilities as 
        \begin{align}
            \label{eq:moctail}
            \mathbb{P}\{x>\nu|x>\mu\}
            \approx
            \widehat{Q}(\nu,\mu; N,D,A)
            :=
            \frac1N\sum_{n=1}^N
            \frac{
                \sum_{d=1}^D\mathbb{I}\{x_{d,n}^*>\nu\}
            }{
                \sum_{d=1}^D\mathbb{I}\{x_{d,n}^*>\mu\}
            }.
        \end{align}
\end{enumerate}

The estimator $\widehat{Q}$ is called the ``mixture of conditional tails'' (MoCTail) estimator, formulated in \cite{Finkel2026boosting} and similar to but distinct from the estimator of \cite{BloinWibe2025estimating}. The arguments before the semi-colon ($\nu,\mu$) specify what the tail region is, and those after the semi-colon ($N,D,A$) are algorithmic parameters. We hope for $\widehat{Q}$ to approximate the true climatological tail distribution, which can be estimated by DS: that is, running an exremely long simulation of length $T$ in which a growing number $N(T)\propto T$ events appear,
\begin{align}
    \mathbb{P}\{x>\nu|x>\mu\}=Q(\nu,\mu)=\lim_{T\to\infty}\frac1{N(T)}\sum_{n=1}^{N(T)}\mathbb{I}\{x_n^*>\nu\},
\end{align}
where each $x_n^*$ is a peak over $\mu$. But the estimator's fundamental validity, let alone its efficiency with small $N$, is not proven, and might depend sensitively on several hyperparameters of the boosting procedure, most notably the advance split time $A$ and the form of perturbation $\delta x$. The optimal choice may well be different between families. 

The next section makes the question more concrete by focusing on the Logistic Map, but along the way take a detour to the Tent Map. The pair of systems together clarify some conditions on $\delta x$ and $A$ for EB to be valid and efficient. 

\section{Ensemble Boosting on a minimal system}

\subsection{Logistic Map}
\label{sec:logisticmap}

\begin{figure}
    \centering
    \includegraphics[width=0.99\linewidth,trim={0cm 0cm 12cm 0cm},clip]{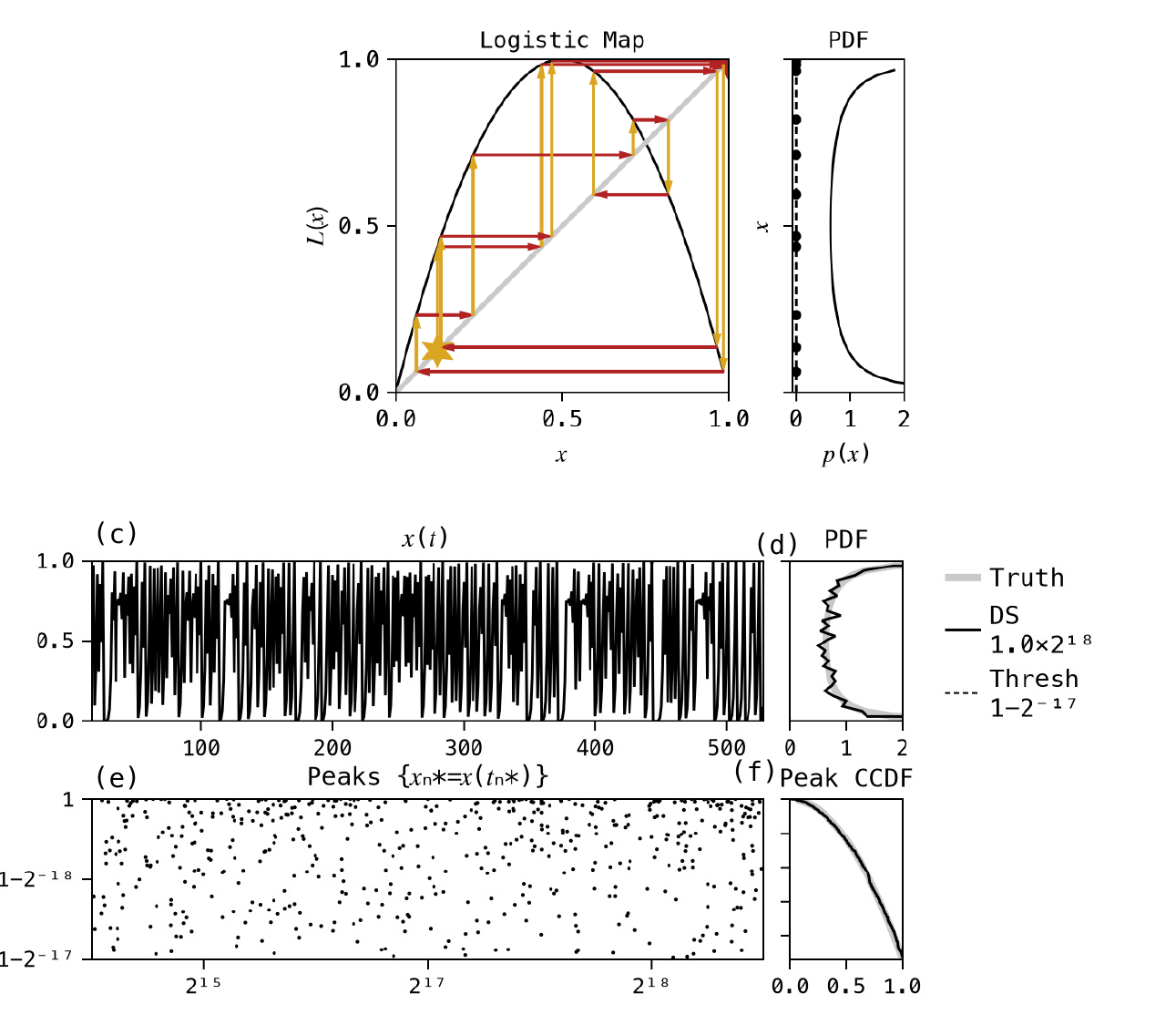}
    \caption{Illustration of the Logistic Map. (a) Eleven iterations plotted by the visual method: starting from the yellow star, we move vertically to the graph $(x,L(x))$ to advance the system to the next timestep, and then horizontally to the diagonal line $(x,x)$ to move into position for the next advancement. The zig-zag nearly closes in a loop after 9 iterates, but the small gap already starts to amplify by the 11th iterate. This is the hallmark of chaos. (b) Those same iterates arranged in a line next to the limiting probability density function (PDF) which they are samples from. (c) A timeseries shows sustained irregular, aperiodic behavior. (d) The empirical PDF from a longer timeseries (duration $2^{18}$) overlaid on the true PDF to which it would converge given enough time. (e) Exceedances over the threshold $\mu\approx1-2^{-17}$, and (f) the complementary cumulative distribution function (CCDF) of peaks, i.e., $Q(\nu,\mu)=\mathbb{P}\{x>\nu|x>\mu\}$ conditional on already exceeding the threshold $\mu$. The goal of RES is to efficiently evaluate this CCDF of peaks. } \label{fig:illustration_logisticmap}
\end{figure}

The essential extreme events and the need for EB are already seen in the Logistic Map, $F(x)=L(x)=4x(1-x)$, a common introductory textbook example of chaos \citep{Ott2002chaos}. Fig.\ref{fig:illustration_logisticmap} illustrates its behavior, two aspects of which are central for our aims. (i) There is a steady-state PDF with highly compressed tails at $x\to0$ and $x\to1$, and (ii) trajectories with very similar initial conditions diverge rapidly. The sequence plotted almost repeats itself after 9 steps, but then strays quickly from its past history. EB relies on this error growth to transform tiny perturbations $\{\delta x(t^*-A)\}$ into major differences in event severity ($\delta x(t^*)$). 

It seems clear that the right choice of AST to boost a peak $x^*$ should depend on how fast errors grow leading up to $x^*$. The one-step growth of a small initial error $\delta x_0$ is proportional to the local slope of $F(x)$, and after $t$ steps it compounds into
\begin{align}
    \delta x(t)
    &\approx
    \frac{dx(t)}{dx(t-1)}\frac{dx(t-1)}{dx(t-2)}\hdots\frac{dx(1)}{dx(0)}\delta x(0) \\
    &=F'(x(t-1))F'(x(t-2))\hdots F'(x(0))\delta x(0) \\
    &=\pm2^{t\lambda(x(0);t)}\delta x(0) \\
    \text{where }
    \lambda(x(0);t)&:=\frac1t\log_2\big|F'(x(t-1))\hdots F'(x(0))\big|=\frac1t\lim_{|\delta x(0)|\to0}\log_2\bigg|\frac{\delta x(t)}{\delta x(0}\bigg|.
    \label{eq:ftle_defn}
\end{align}
The finite-time Lyapunov exponent (FTLE) $\lambda(x(0),t)$ quantifies the error growth rate over the fixed time horizon $t$ as an inverse doubling time (for convenience we use base 2 rather than the more standard base $e$, and all logarithms written henceforth are base 2 by default). It obviously depends in a complicated way on the precise location of $x^*$, being a joint function of the next $t$ subsequent iterates. So already the choice of $A$ is complicated, even in the humble Logistic Map. But fortunately, following the textbook analysis presented in \cite{Ott2002chaos}, we can transform the system to one where all slopes are equal, which will yield an exact form of the steady-state PDF and the optimal AST for approximating it.

\subsection{Transformation to the Tent Map}
A change of variable $x=\sin^2(\frac12\pi z)=:h^{-1}(z)$ induces an equivalent dynamics on $z=\frac2\pi\sin^{-1}(\sqrt{x})=:h(x)$:
\begin{align}
    x(t+1) 
    &= 4x(t)(1-x(t)) \\ 
    \sin^2\Big(\frac12\pi z(t+1)\Big)
    &= 4\sin^2\Big(\frac12\pi z(t)\Big)\Big[1-\sin^2\Big(\frac12\pi z(t)\Big)\Big] \\
    &= \Big[2\sin\Big(\frac12\pi z(t)\Big)\cos\Big(\frac12\pi z(t)\Big)\Big]^2 \\
    &= \sin^2\Big(\frac12\pi\cdot2z(t)\Big) \\
    &= \sin^2\Big[\frac12\pi\cdot2\big(1-z(t)\big)\Big]  && (\sin^2\text{ is even and $\pi$-periodic}) \\ 
    &= \sin^2\Big[\frac12\pi\cdot2\min\{z(t),1-z(t)\}\Big] && (\text{combine previous two expressions})\\
    \therefore z(t+1) &= 2\min\{z(t),1-z(t)\}=:T(z(t)) && \text{Tent Map} 
\end{align}

\begin{figure}
    \centering
    \includegraphics[width=0.99\linewidth,trim={0cm 0cm 8cm 0cm},clip]{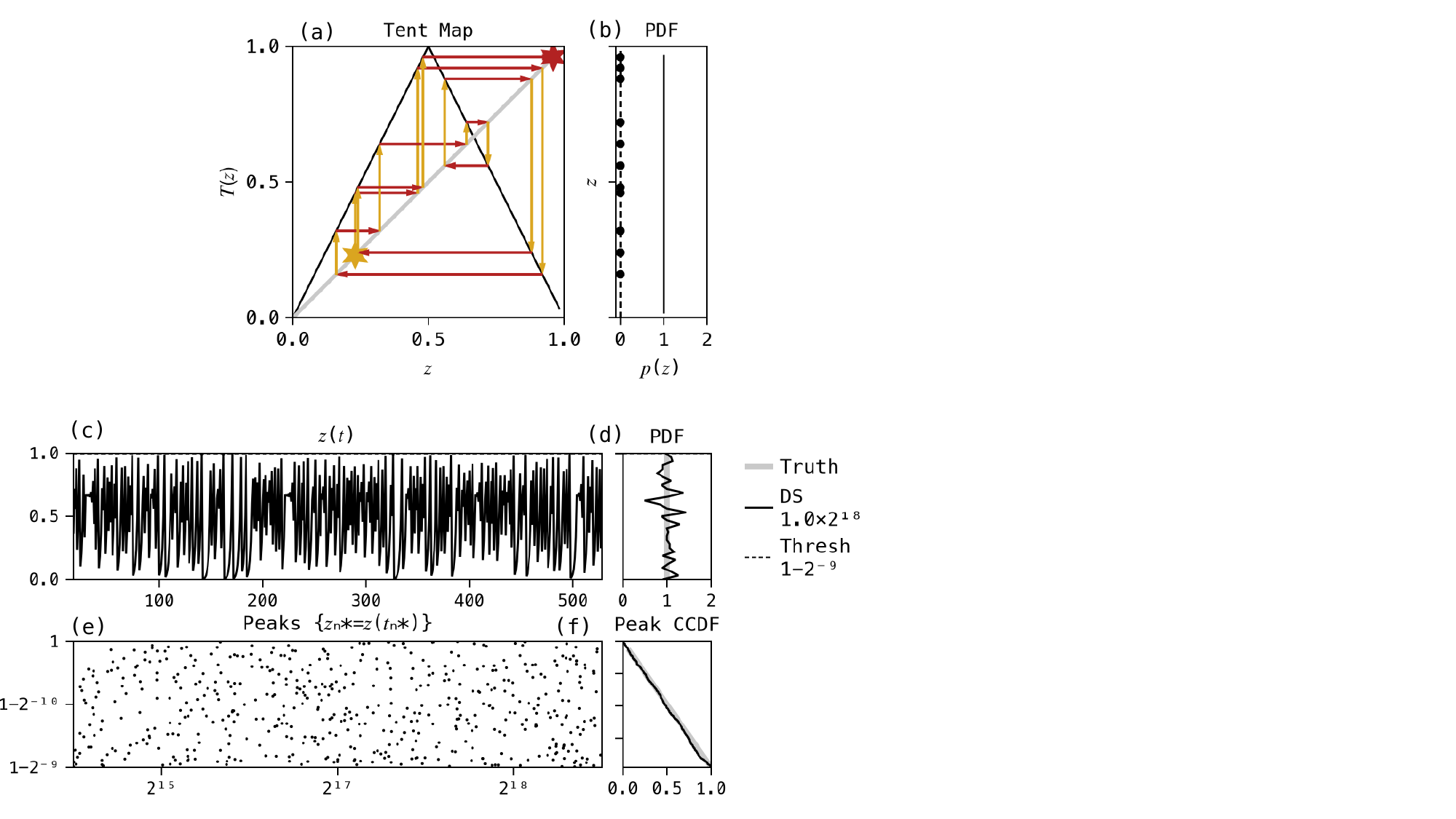}
    \caption{Illustration of the Tent Map, formatted the same as Fig.\ \ref{fig:illustration_logisticmap}}.
    \label{fig:illustration_tentmap}
\end{figure}

The last line of algebra is needed so that $z(t+1)$ remains in $(0,1)$. This is the well known \emph{conjugacy mapping} $h:x\mapsto z$, which preserves topological properties like fixed points and periodic orbits, but warps the geometry. The Tent Map and its associated dynamics is illustrated in Fig.\ \ref{fig:illustration_tentmap}. The ten-step trajectory looks the same topologically as in Fig.\ \ref{fig:illustration_logisticmap}, but doesn't go so close to the edges of $(0,1)$. The steady-state PDF for $T(z)$ is in fact \emph{exactly uniform}: if $z(t)$ (the weather today) is a uniform random variable over $(0,1)$, its possible states can be visualized as a regularly spaced point cloud. After one timestep, the half of the cloud below $1/2$ will spread evenly along the vertical axis tomorrow---only with half its original density---and so will the subset of points above $1/2$, making tomorrow's weather exactly uniformly spread out just like today's. (Contrast this with the Logistic Map $L(x)$, which concentrates probability close to the extremes by sending trajectories readily towards $x=1$ and pulling them reluctantly away from $x=0$.) The PDF for the Logistic Map can then be calculated simply from the change-of-variable formula:

\begin{align}
    p_x(x) &= \frac{p_z(z)}{|dx/dz|} 
    =\frac{1}{2\sin(\frac12\pi z)\cos(\frac12\pi z)\cdot\frac12\pi}
    =\frac1{\pi\sqrt{x(1-x)}},
    \label{eq:logisticmap_pdf}
\end{align}
which confirms the high-density boundaries ($p(x)\to\infty$ as $x\to0$ and $x\to1$). 

Representing $z(t)$ in its binary expansion, the update rule can be reduced to operations on its bits:
\begin{align}
    z(t) 
    &= 0.\beta_1\beta_2\beta_3\beta_4\hdots \\
    z(t+1)
    &= 
    \begin{cases}
        0.\beta_2\beta_3\beta_4\beta_5\hdots & \text{ if }\beta_1=0 \\ 
        0.\bbar_2\bbar_3\bbar_4\bbar_5\hdots & \text{ if $\beta_1=1$ (where $\bbar=1-\beta$)}
    \end{cases}
\end{align}
A computer with finite memory can only represent so many bits---in our examples, we compute with 32-bit integer arithmetic---and so has to decide what happens to the 32nd bit after each timestep. We set it randomly from an unbiased coin flip.

\subsection{Ensemble Boosting on the Tent Map}
\label{sec:ebtent} 
We now walk through the EB steps from Section \ref{sec:ebwhite} on the Tent Map, presenting empirical results while explicitly tracking bits to derive the theoretical prescription for optimal AST. 

\begin{centering}
    \begin{figure}
        \includegraphics[width=0.9\linewidth,trim={0cm 0cm 17cm 0cm},clip]{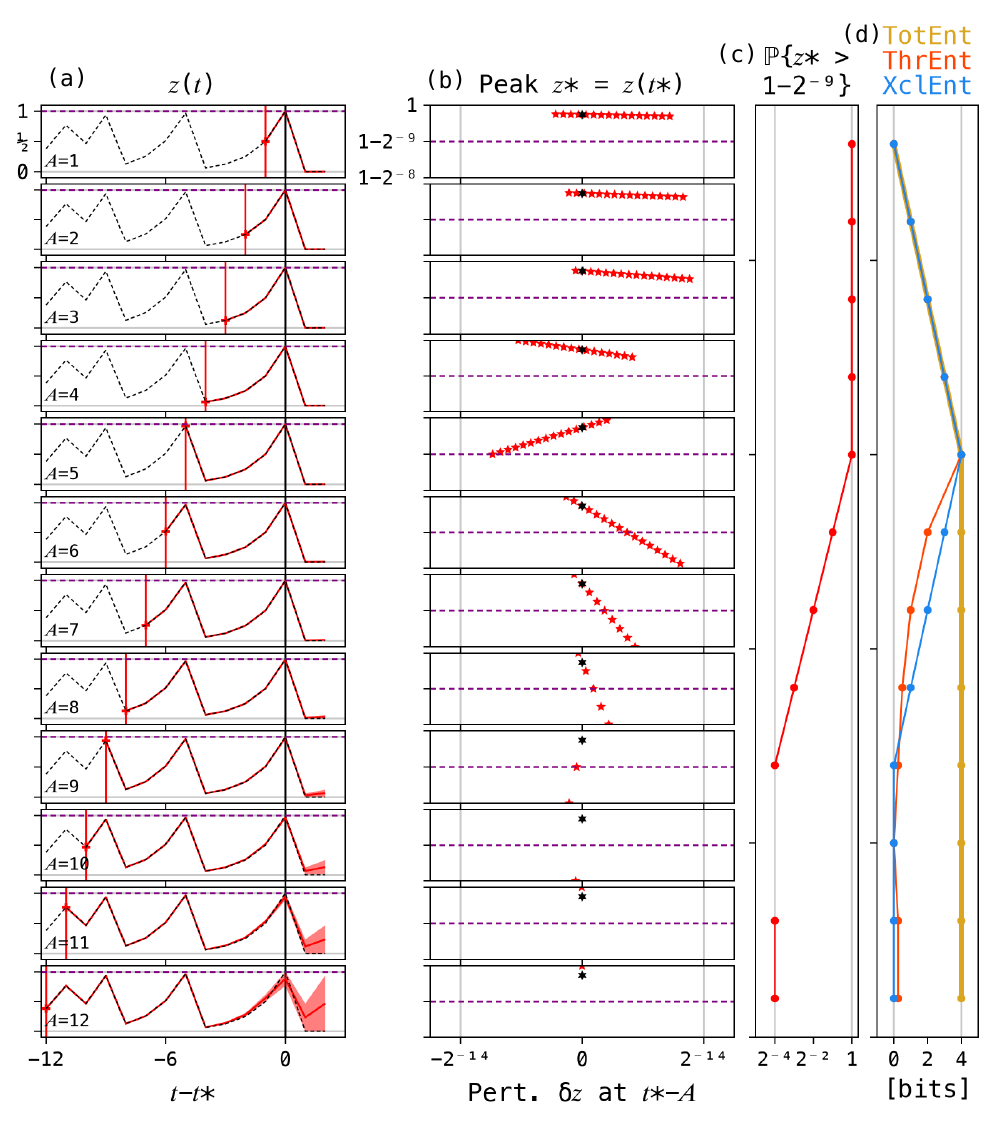}
        \caption{Ensemble Boosting of a single ancestor with the Tent Map, each row showing a different AST from $A=1$ to $A=12$. In this example, the perturbation magnitudes are $\delta z=2^{-K}=2^{-14}$, and the threshold defining ``extreme'' is $\mu=1-2^{-M}=1-2^{-9}$. The descendant ensemble size is $D=16$ at each AST. (a) An ancestor event (black timeseries) peaks at $t^*$ (black guideline) above $\mu$ (horizontal dashed line, visually indistinguishable from 1). A pool of descendants is launched at $t^*-A$ (red guideline). Red timeseries show the ensemble mean and the min-max range across all $D=16$ descendants at each AST. (b) Scatterplots showing peaks $z^*$ at time $t^*$ as a function of the perturbations $\delta z$ applied earlier at $t-t^*$. Since $\delta z$ is a bit truncation followed by a bit-flipping, it is not always symmetric about zero. The ancestor's peak, in black, sits above the threshold $\mu$. So do the descendants' peaks for short AST, but for $A>5$ they increasingly fall below $\mu$, as quantified in (c) the fraction of descendant peaks still in the tail. (d) Three forms of entropy quantify the spread of descendant severities; they increase in tandem with $A$ for $A<5$, and subsequently diverge. \totent\ plateaus at 4 bits, the maximum \emph{empirical} entropy achievable by a 16 ($=2^4$)-member descendant ensemble over the $2^{-M}/(\text{bin width})=2^{-9}/2^{-13}=2^4$ bins of the tail. See text for rationale and definitions.} 
        \label{fig:ebtent_storyline}
    \end{figure}
\end{centering}

\begin{centering}
    \begin{figure}
        \includegraphics[width=0.99\linewidth,trim={0cm 0cm 8cm 0cm},clip]{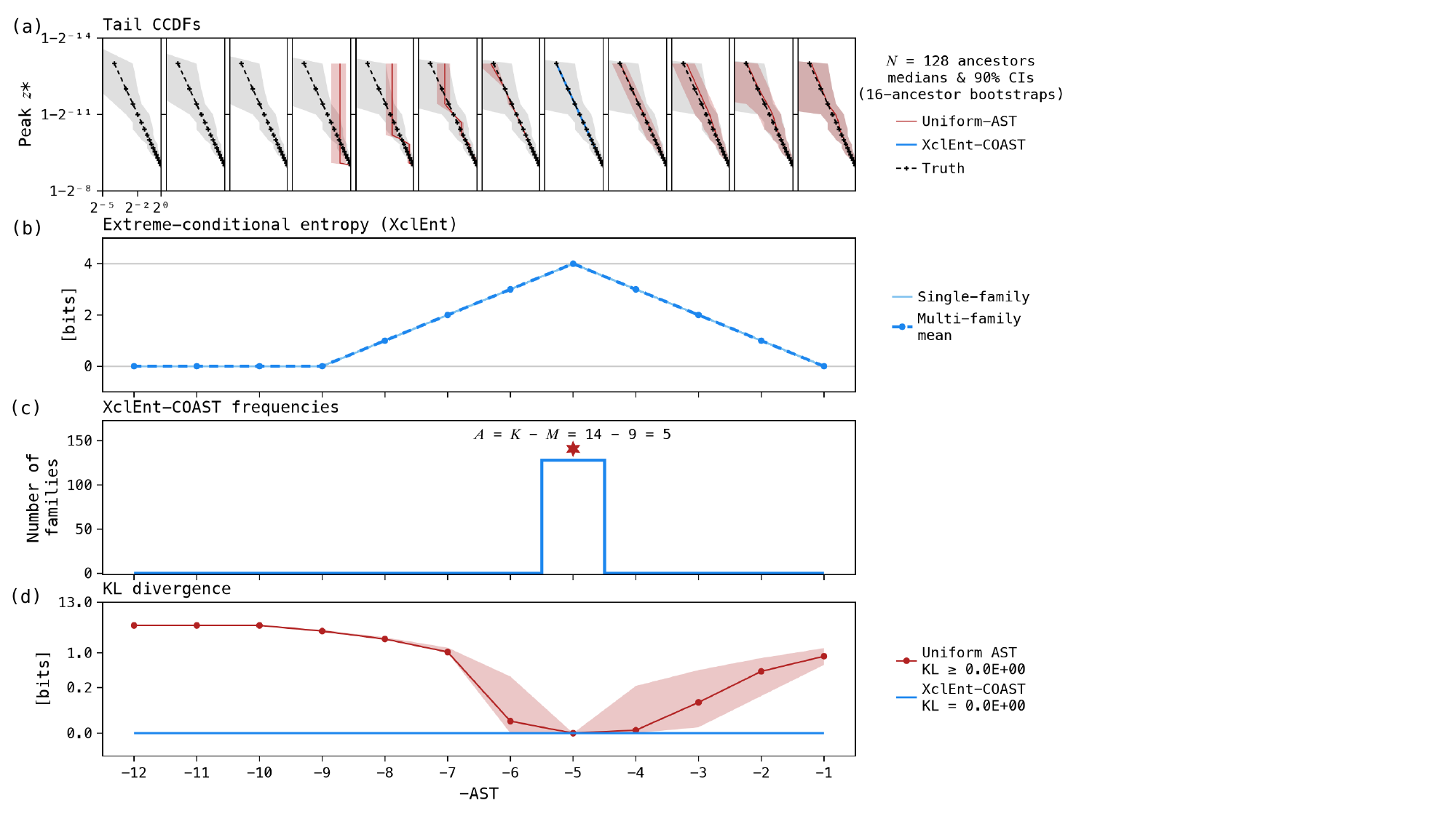}
        \caption{
            Performance of EB for the Tent Map, from the same experiment as in Fig.\ \ref{fig:ebtent_storyline} but now aggregated over ancestors. 
            (a) Tail CCDFs according to (gray) empirical CCDFs of ancestor peaks, without boosting; (red) MoCTail estimator from Eq. (\ref{eq:moctail}), where a uniform AST is chosen for each ancestor, $A=1$ at the rightmost panel and $A=12$ at the leftmost panel; (blue) MoCTail estimator with $A_n$ chosen to maximize extreme-conditional entropy (\xclent) separately for each ancestor (in this case, $A=5$ is always chosen), and (black) the true tail CCDF. Solid lines and uncertainty bands show the medians and 5-95 percentile ranges for MoCTail estimates using only 16-family subsets of the full 128-family population available. The left three panels ($A\geq9$) have no MoCTails because all descendants have left the tail. The MoCTails at $A=5$ have no uncertainty for the Tent Map, but this will change with the Logistic Map.
            (b) \xclent\ as a function of AST for each family separately (light blue) and \xclent\ averaged across all 128 families (dark blue). Here, for the Tent Map, they are all identical; they will not be for the Logistic Map. 
            (c) Histogram of \xclent-maximizing ASTs across families, i.e., $\sum_{n=1}^N\mathbb{I}\{$family $n$'s \xclent\ is maximized at AST $A\}$, as a function of $A$. Again, this histogram is degenerate for the Tent Map but will have some spread for the Logistic Map. 
            (d) Error in MoCTail estimator measured by KL divergence, as a function of uniformly AST (red) or if each family maximizes \xclent\ separately (blue), with error bands made the same way as in (a). Error bands are zero for the \xclent-\coast\ rule here, but will have some width for the Logistic Map. 
        }
        \label{fig:ebtent_statistics_K14_M9}
    \end{figure}
\end{centering}

\begin{enumerate}
    \item From a DS run of duration $2^{16}$, we identify peaks $\{z_n^*=z(t_n^*)\}_{n=1}^N$ above the threshold $\mu=1-2^{-9}=:1-2^{-M}$, as shown in Fig.\ \ref{fig:illustration_tentmap}. Because $z$ has a uniform distribution, any random snapshot has a $2^{-9}$ chance of exceeding $\mu$, so we expect to find $2^{16-9}=128$ exceedances. For this particular system, every exceedance is an isolated peak (no declustering needed), because (provided $\mu>\frac23$) both the preceding and following states $z(t^*\pm1)<\frac23$. Enforcing a minimum time delay of $t_{n+1}^*-t_n^*\geq2^4=16$ between consecutive peaks to ensure statistical independence (16 is an ample buffer, since it only takes 9 consecutive bits to define a peak) removes a negligible few, leaving exactly $N=128$ peaks. Independence is not strictly required, but makes statistical estimates more stable. Fig.\ \ref{fig:ebtent_storyline}a shows the DS snippet leading to the first peak.

    \item 
        Identify the antecedent states $z(t^*-A)$, marked in Fig.\ \ref{fig:ebtent_storyline}a by vertical red lines, for $A$ ranging from 1 to 12 for going down the column. Since the peak $z(t^*)>1-2^{-M}$, its binary expansion starts with $M$ leading 1s, 
        \begin{align}
            z(t^*)=z^*
            &=
            0.\underbrace{111\hdots1}_M\beta_{M+1}\beta_{M+2}\beta_{M+3}\beta_{M+4}\hdots,
        \end{align}
        and winding back the clock by $A$ time units has the effect of inserting $A$ bits in front extracted from the timeseries:
        \begin{align}
            z(t^*-A)
            &=0.\alpha_1\hdots\alpha_A
            \begin{cases}
                \overbrace{11\hdots1}^M\beta_{M+1}\beta_{M+2}\beta_{M+3}\beta_{M+4}\hdots & \text{if }\alpha_1+\hdots+\alpha_A\text{ is even} \\
                \underbrace{00\hdots0}_M\bbar_{M+1}\bbar_{M+2}\bbar_{M+3}\bbar_{M+4}\hdots & \text{if }\alpha_1+\hdots+\alpha_A\text{ is odd}. 
            \end{cases}
        \end{align}
        Note that \emph{any} interval on $(0,1)$ with length $2^{-M}$ could be used as the definition of ``extreme'', by prescribing some other sequence of $M$ bits instead of $1\hdots1$, and the same logic would apply. Incidentally, this point demonstrates that EB is highly customizable, a major practical asset because of the complicated ways that physical weather hazards can propagate through ecosystems. Riffing on adages, one might say that \emph{extremity is in the eyes of the stakeholder}, who in this example can define the interval of interest by nailing two tent stakes in the ground a distance $2^{-M}$ apart, anywhere within the interval $(0,1)$. \emph{If all you hold is a stake (hammer), anything can look like a tail.}

    \item 
        Perturb this antecedent state by a random number $\delta z$ with magnitude $\leq2^{-14}=:2^{-K}$, randomly flipping each bit past the $K$th independently with probability $\frac12$ each. The new perturbed state can be written as
        \begin{align}
            z(t^*-A)+\delta z
            &=0.\alpha_1\hdots\alpha_A
            \begin{cases}
                \overbrace{11\hdots1}^M\beta_{M+1}\beta_{M+2}\hdots\beta_{K-A}\gamma_1\gamma_2\hdots & \text{if }\alpha_1+\hdots+\alpha_A\text{ is even} \\
                \underbrace{00\hdots0}_M\bbar_{M+1}\bbar_{M+2}\hdots\bbar_{K-A}\gbar_1\gbar_2\hdots & \text{if }\alpha_1+\hdots+\alpha_A\text{ is odd},
            \end{cases}
        \end{align}
        where each $\gamma\in\{0,1\}$ is a random bit. By writing out all of the $\alpha$s, 1s/0s, and $\beta$s/$\overline{\beta}$s, we've assumed $K>A+M$, but in general, there will be $\min\{K,A\}$ leading $\alpha$s, followed by $\max\{0,\min\{K-A,M\}\}$ 0s or 1s, and then $\max\{0,K-A-M\}$ $\beta$s or $\overline{\beta}$s before the random bits start. These three cases are pivotal to consider in the coming argument. 

    \item 
        Evolve the perturbed state forward again, arriving after $A$ timesteps at the perturbed descendant peak
        \begin{align}
            \label{eq:peak_bits_desc}
            T^A(z(t^*-A)+\delta z)
            &=
            0.\overbrace{11\hdots\hdots\hdots\hdots\hdots1}^{
                \max\{0,\min\{K-A,M\}\}    
            }
            \underbrace{\beta_{M+1}\beta_{M+2}\hdots\beta_{K-A}}_{
                \max\{0,K-A-M\}
            }
            \gamma_1\gamma_2\hdots.
        \end{align}

    \item 
        Repeat steps 2-4 with perturbations $\{\delta z_d\}_{d=1}^D$ for $D=16$ descendants from each AST. Fig.\ \ref{fig:ebtent_storyline}b plots the perturbed peaks (red) as a function of the perturbation $\delta z$ taking place $A$ steps earlier, for each $A$. We don't actually choose $\delta z$s randomly, but quasi-randomly, with the \emph{low-discrepancy} van der Corput sequence on the unit interval, with noted favorable properties for sampling and stochastic computing \citep{Moghadam2024accurate}:
        \begin{align}
            \{v_1,\hdots,v_D\}
            &=\bigg\{
                0,\frac12,
                \frac14,\frac34, 
                \frac18,\frac38,\frac58,\frac78,
                \frac1{16},\frac3{16},\hdots,\frac{15}{16} 
             \bigg\} \notag\\
             &=\Big\{
                 0.0, 0.1, 
                 0.01, 0.11, 
                 0.001, 0.011, 0.101, 0.111, 
                 0.0001, 0.0011, \hdots, 0.1111
             \Big\},
         \label{eq:vandercorput}
        \end{align}
        and using the corresponding binary expansion to overwrite the $(K+1)$th-and-following bits of $z(t^*-A)$. In other words, 
        \begin{align}
            z(t^*-A)+\delta z_d=2^{-K}\Big(v_d+\Big\lfloor2^Kz(t^*-A)\Big\rfloor\Big)
            \label{eq:dzimplicit}
        \end{align}
        This defines the $\delta z_d$s implicitly from $v_d$s. The ensemble of $\delta z_d$s may not be symmetric about zero, depending on $z(t^*-A)$, which explains why the red point clouds in Fig.\ \ref{fig:ebtent_storyline}b are lopsided left or right at each lead time, though on average they are symmetric. The advantage of this quasi-random perturbation sequence is better estimation of statistics like the MoCTail estimator, at least in low dimensions, and this particular sequence is easily extensible in case the initial descendant pool $D$ is deemed too small. 
        Fig.\ \ref{fig:ebtent_storyline} shows how the peak's height depends jointly on $\delta z$ and $A$. At short $A$, $\delta z$ has hardly any effect, and all the descendant peaks sit next to the ancestor, safely above the threshold $\mu$. $\delta z$ exerts a stronger influence at longer $A$, enough to knock some descendant peaks below $\mu$. Column (c) shows the fraction of descendants still exceeding $\mu$ gets cut in half repeatedly with each increment of $A$ from 5 to 10, after which all 16 remain below $\mu$ except an occasional fluctuation. A bit of reasoning explains this behavior. A descendant peak exceeds $1-2^{-M}$ if its binary expansion begins with at least $M$ 1s, requiring that perturbations are small enough not to interfere with the first $A+M$ bits in the descendant's initial conditions ($A$ timesteps pre-peak): $K\geq A+M$, so $A\leq K-M=14-9=5$, which is exactly the longest AST at which all descendants remain in the tail in Fig.\ \ref{fig:ebtent_storyline}b. 

    \item 
        Repeat step 5 (and by recursion, steps 2-4) for ancestors $n=1,\hdots,N$. Though the ancestor peaks will vary in position from $1-2^{-M}$ to 1, the descendant pool always remains within the tail given an AST $A\leq K-M$, and increasingly leaks out of the tail given longer ASTs.

    \item 
        Estimate the climatological tail using the MoCTail estimator in Eq. (\ref{eq:moctail}). Fig.\ \ref{fig:ebtent_statistics_K14_M9}a displays the resulting MoCTail CCDFs (red) for each AST increasing from right to left, overlaid on the true tail CCDF (black) and the empirical CCDF we would obtain from ancestor peaks alone without boosting (formally equivalent to $A=0$). Each CCDF uses the same bins, whose lower edges are linearly spaced, $2^{-13}$ apart, and marked by black circles (note log scales on both axes). Error bands are 90\% percentile bootstrap ranges, estimated by sub-sampling the 128 families with replacement into groups of size 16 and repeating 500 times. The solid lines are medians over such sub-samples. 

        A good-quality MoCTail estimator is one that matches the truth with small uncertainty, by which criterion Fig.~\ref{fig:ebtent_statistics_K14_M9}a indicates a clear winner of $A=5=K-M$, the same identified in step 5 as the longest AST that still confines all descendants to the tail. The binary representation of descendant peaks in Eq.~(\ref{eq:peak_bits_desc}) explains exactly why: the leading $\max\{0,\min\{K-A,M\}\}$ bits are all 1, the trailing bits from position $K+1$ onward (the $\gamma$s) are random coin flips, and the intermediate $\max\{0,K-A-M\}$ bits are copied from the ancestor. The best possible distribution of descendants is uniform across the tail to match the climatology distribution, which is uniform for the Tent Map. A random number uniformly drawn from the tail has $M$ leading 1s followed by random bits. This is achieved by the descendant ensemble precisely by retaining none of the unnecessary $\beta$s---or, to make a genetic analogy, letting descendant genomes mutate as much as possible while retaining some minimal shared ancestry---thus setting $K-A-M=0$. 

\end{enumerate}

To summarize, we have deduced a simple, transparent joint relationship between the size $2^{-M}$ of the target set, the maximum perturbation size $2^{-K}$, and the optimal advance split time $A$:
\begin{align}
    \label{eq:coastformula}
    A=K-M.
\end{align}

In one sense, we have just ``solved'' the EB optimization problem for the Logistic Map, as well as all other maps conjugate to the Tent Map. The recipe is to (1) transform $x$ to $z$ by the conjugacy $x=\sin^2(\frac12\pi z)$; (2) perform EB in $z$-space by applying perturbations $\delta z$ given by Eqs.\ (\ref{eq:vandercorput}-\ref{eq:dzimplicit}) with magnitude $2^{-K}$ at $A=K-M$ timeseps before each peak; and (3) translate the resulting tail estimates of interest back into $x$-space. This procedure amounts to adjusting the perturbation $\delta x$ according the initial condition, $\delta x=\big|\frac{dx}{dz}\big|\delta z=\pi\sqrt{x(1-x)}\delta z$, while keeping the AST fixed at $K-M$. Note that $|\delta x|\to0$ as $x\to\pm1$, compensating for faster error growth with smaller perturbations. If a practitioner can invent such a transformation, this procedure is elegant and effective. But in general, optimizing perturbation magnitude and shape can be a daunting search problem in a vast high-dimensional landscape. This is why our recent and present focus is on perturbation timing only, assuming a fixed perturbation shape. The next section lays out a principle for attacking this simpler, one-dimensional optimization problem, to get the best possible performance given a possibly restrictive constraint. 

\subsection{Interpreting the optimal advance split time}
\label{sec:interpreting}

The most intuitive way to read Eq.~(\ref{eq:coastformula}) is $2^{-K}2^A=2^{-M}$: A bubble of perturbations that starts with small volume $2^{-K}$ will expand, over the course of $A$ timesteps, to a size that matches the target set $(1-2^{-M},1)$. Another way to read it is in reverse-time, $2^{-M}2^{-A}=2^{-K}$: a cloud of points spread uniformly across the tail at time $t^*$, with volume $2^{-M}$, is mapped under the \emph{backward} dynamics $T^{-1}(\{z\})=\{\frac{z}{2},1-\frac{z}{2}\}$ for $A$ timesteps into a scattered collection of $2^A$ possibly-overlapping intervals, each receiving an equal share of the original volume. Thus, $z(t^*-A)$ sits somewhere in the preimage $T^{-A}((1-2^{-M},1))$ on an interval with length at least $2^{-M}2^{-A}$, dictating how big the perturbation can be. 

In the current rare-event-sampling literature, there are several competing heuristics for optimal AST, which in light of Eq.~(\ref{eq:coastformula}) can quickly be dispelled as not nuanced enough. 
\begin{itemize}
    \item 
        One heuristic is to use the Lyapunov (or error-doubling) timescale, which for the Tent Map is $1$. Simply setting $A=1$ would be obviously wrong, but even setting $A$ as \emph{any} constant multiple of the Lyapunov time---a value intrinsic to the dynamical system---cannot respect the need for $A$ to vary as a function of $K$ and $M$. This simplistic criterion is not, to our knowledge, actually promoted in any published studies, but it is the most common first instinct in casual discussions. 

    \item 
        Another heuristic is to use the time $t_\epsilon$ needed for perturbations to saturate to a fraction $\epsilon$ of their long-time size \citep{Finkel2024bringing,Finkel2026rare}. For the Tent Map, the initial error is $2^{-K}$ and the long-time error is 1, so $2^{-K}2^{t_\epsilon}=\epsilon\implies t_\epsilon=K+\log\epsilon$. This AST prescription now does vary with $K$, but $\epsilon$ is a mysterious hyperparameter that has to be tuned to the system. Our new formula suggests that $\epsilon=2^{-M}$. Here, that is simply the overall probability of the tail $\{z\geq\mu\}$, but extrapolating this interpretation would be a mistake. In preceding work, we estimated $\epsilon=\frac38$ in the Lorenz-96 system \citep{Finkel2024bringing} and $\epsilon=\frac23$ in an aquaplanet atmospheric model \citep{Finkel2026rare}, in both cases targeting probabilities $\sim1/1000$. We suspect the discrepancy comes from the coexistence of \emph{many} timescales in these spatially extended systems, and that the optimal AST lies well beyond the regime of small, exponentially growing errors in which Lyapunov analysis applies. This same point is recognized by \cite{BloinWibe2025estimating}. The Tent Map, being piecewise linear, maintains a steady error growth rate all the way to saturation. In general, $A$ might be better interpreted as a \emph{finite-size} Lyapunov timescale \citep{Boffetta1998extension}, which would incorporate dependence on $K$. But relating it to $M$ might still be tricky; for example, $\log\epsilon$ might be interpreted as either the extreme event set's diameter, volume, or some other notion of measure. 
    \item 
        Another idea for optimal AST comes from \cite{BloinWibe2025estimating}, who propose to maximize some fixed upper quantile of the descendant distribution. They compute an explicit formula for AST (what they call ``optimal lead time'') on a linear stochastic Gaussian (Ornstein-Uhlenbeck) process with given restoring force and diffusion coefficient, and then translate it into practice by estimating equivalent parameters in a climate model. They go further to aggregate over multiple ASTs. This sensible and pragmatic strategy accounts for the need to go beyond infinitesimal errors, and incorporates dependence on stochastic perturbation strength (the diffusion coefficient) as well as on the user-defined extreme event set of interest. However, the latter dependence is encoded very indirectly through the choice of upper quantile (their parameter $k$, which also appears in reinforcement learning by the name ``Upper Confidence Bound'' \citep{Cappe2013kullback}). Furthermore, their optimal lead time actually increases with noise strength (their Eq. B5), a totally opposite result from what we found with the Lorenz-96 model \citep{Finkel2024bringing}. We attribute the discrepancy to the Ornstein-Uhlenbeck's fundamentally non-chaotic, noise-driven character: longer AST gives the noise more time to drive the process into the tail, a very different mechanism from chaotic error growth, and this might limit how much that result can be applied to chaotic systems. 

\end{itemize}

 We suspect that any effort to generalize Eq.\ (\ref{eq:coastformula}) for climate models, with all their added complexity,  will have to draw on ergodic theory, fractal geometry, and extreme-value statistics, building on the connections found in \cite{Lucarini2014towards,Lucarini2016extremes} and related work. We consider it a high priority to advance theoretical research along these lines, coupled with systematic numerical benchmarking. 

For lack of a comprehensive theory at present, we defer inspired speculation and re-center the quantitative end goal: to choose an AST that lets the descendant ensembles achieve, collectively through the MoCTail estimator, the best possible approximation of the true climatological tail. In principle, this could be measured by an arbitrary loss function dictated by stakeholder concerns, for example thresholds (like the height of a dam, bridge, or floodwall) where monetary or physical cost jumps. In our current study, we start by making a specific choice of loss function, discussing its ramifications later in Section \ref{sec:discussion}. 

We choose to quantify tail estimation error by Kullback-Leibler (KL) divergence. Given two probability mass functions (PMFs) $\bq=\{q_b\}$ and $\bp=\{p_b\}$ over a set of bins $b$,
\begin{align}
    \text{KL divergence of $\bp$ from $\bq$}=\kl{\bp}{\bq}&=\sum_bp_b\log\bigg(\frac{p_b}{q_b}\bigg).
\end{align}
This is a common penalty function used in statistical inference, where one sees random samples from a true distribution $\bp$ and tries to construct an approximation $\bq$. In that setting, $\kl{\bp}{\bq}$ is the expected log-odds ratio that the data came from $\bp$ rather than $\bq$ \citep[e.g.,][ch. 9]{Wasserman2004all}. It is never negative, since $\bp$ did in fact produce the data, and the best $\bq$ can possibly do is to mimic $\bp$ exactly and thus achieve $\kl{\bp}{\bq}=0$.  

We will actually reverse the convention and treat $\bq$ (the second argument) as the climatological truth and $\bp$ (the first argument) as the MoCTail approximation in Eq. (\ref{eq:moctail}), for the practical reason of keeping the KL divergence finite. The MoCTail estimator from EB does not always cover every bin in the tail, especially in a resource-limited setting with small $N$ or $D$, which means some bins may have $p_b=0$ while $q_b\neq0$. Such bins have $p_b\log(p_b/q_b)=0\log0=0$ by convention, but $q_b\log(q_b/p_b)=+\infty$, which is not very informative. On the other hand, if $q_b=0$, the bin is completely off the dynamical attractor, which means $p_b=0$ as well (unless perturbations are unphysical), so the KL divergence is never infinite if we use $\bq$ as climatological truth and $\bp$ as the approximating MoCTail.

To measure the MoCTail's performance on the Tent Map, we divide the tail into bins of uniform width $2^{-13}$, with boundaries $\mu=\nu_0<\nu_1<\hdots<\nu_{B-1}<\nu_B=1$, where $B=\frac{1-\mu}{2^{-13}}=2^{-9+13}=16$ is the number of tail bins. (The number and spacing of bins are yet more design choices which will reveal their arbitrariness more starkly in the next two sections, and which we confront more generally in Section \ref{sec:discussion}.) Fig.\ \ref{fig:ebtent_statistics_K14_M9}d compares the performance of different ASTs in EB for the Tent Map by plotting the KL divergence from the truth of the resulting MoCTail estimator at each $A$. Denote the true PMF as $q_b=Q(\nu_{b-1},\mu)-Q(\nu_b,\mu)$, and denote the MoCTail PMF for each $A$ as $\qhat_b=\widehat{Q}(\nu_{b-1},\mu;N,D,A)-\widehat{Q}(\nu_b,\mu;N,D,A)$ . Then
\begin{align}
    \kl{\widehat{\mathbf{q}}}{\mathbf{q}}=\sum_{b=1}^B\widehat{q}_b\log\bigg(\frac{\widehat{q}_b}{q_b}\bigg)
\end{align}
is the objective function. The best possible value is achieved if $\bqhat=\bq$, making $\kl{\bqhat}{\bq}=0$. The worst possible value is achieved if one bin gets all the probability mass, making $\kl{\bqhat}{\bq}=1\cdot\log(1/2^{-4})=4$. The KL divergence plot in Fig.\ \ref{fig:ebtent_statistics_K14_M9}d confirms that $A=K-M=5$ is the best-performing AST.

\subsection{Generalizing the optimal advance split time with a variational principle}
\label{sec:variational}

This is a happy ending to one story that played out in a minimalist sandbox, but is only the beginning of a longer story of nuances and complications as we step out of the sandbox and move toward generalizing the result to advance practical RES with realistic models. 

\begin{centering}
    \begin{figure}
        \includegraphics[width=0.99\linewidth,trim={0cm 0cm 8cm 0cm},clip]{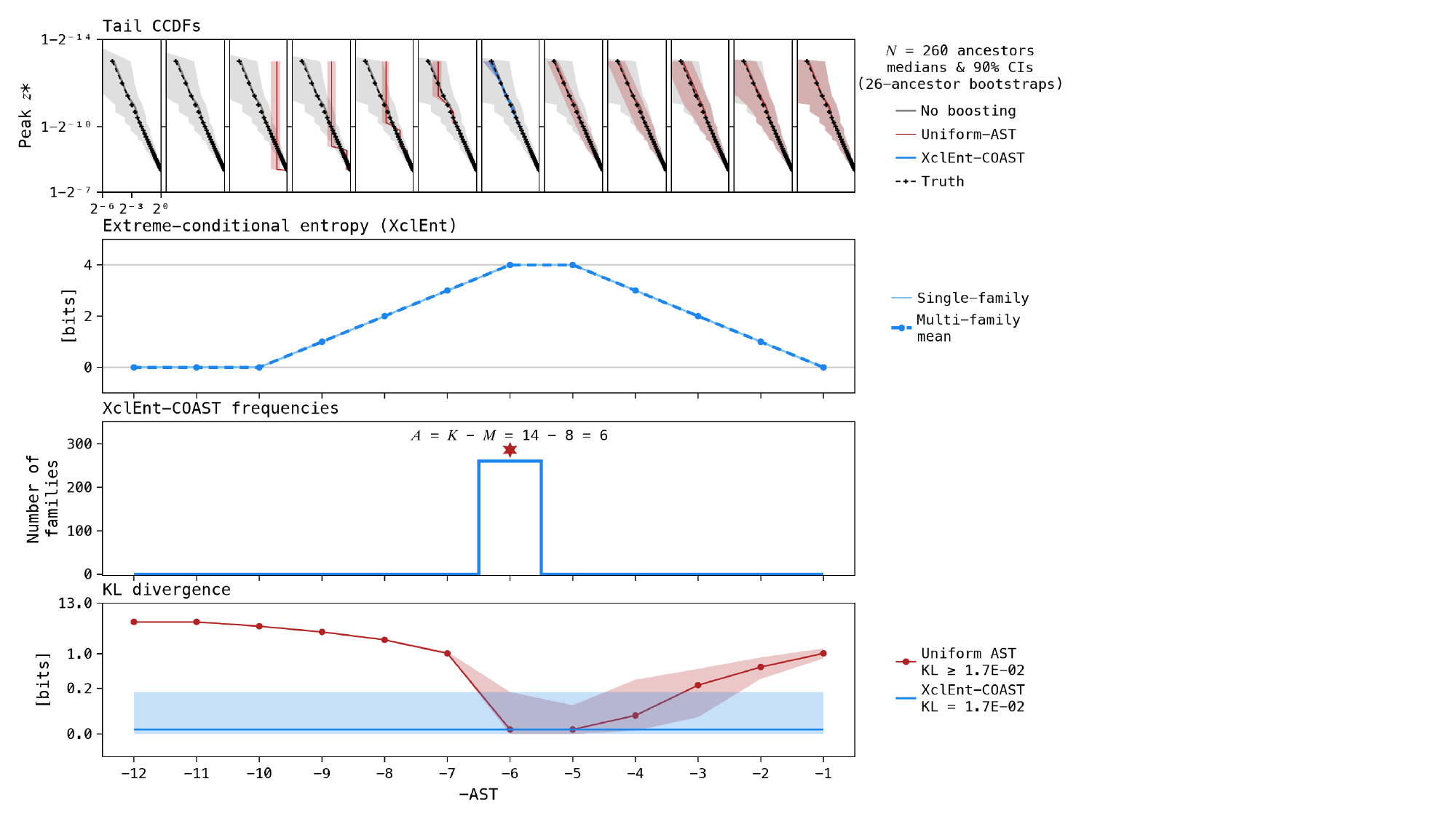}
        \caption{Same as Fig.\ \ref{fig:ebtent_statistics_K14_M9}, but with $(K,M)=(14,8)$.}
        \label{fig:ebtent_statistics_K14_M8}
    \end{figure}
\end{centering}

The fact that $A=5$ achieves the \emph{theoretical minimum} of zero in the preceding example is a hint that our setup so far is too good to be true in general. The first imperfection arises when the bins outnumber the descendants per ancestor, $B>D$, so that any single family cannot possibly cover every bin. Fig.\ \ref{fig:ebtent_statistics_K14_M8} shows MoCTail performance using the same perturbation amplitude ($K=14$) but with a tail twice as long ($M=8$) that includes half-as-rare events of probability $2^{-8}$ rather than $2^{-9}$. The same bin spacing puts $2^{13-8}=32$ bins in the tail instead of 16, so any single 16-descendant pool can only reach half of them. 

\begin{centering}
    \begin{figure}
        \includegraphics[width=0.99\linewidth,trim={0cm 0cm 8cm 0cm},clip]{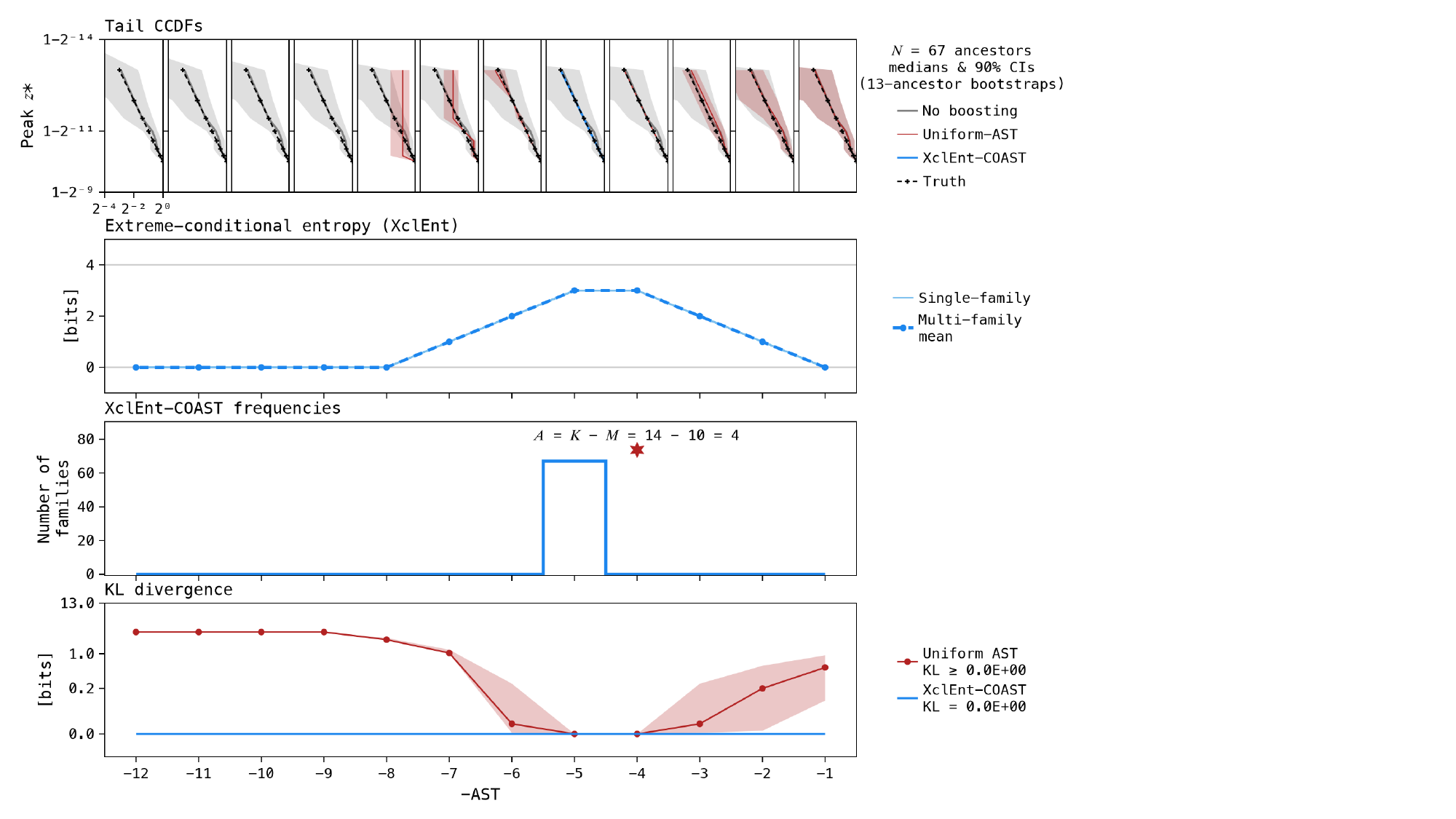}
        \caption{Same as Fig.\ \ref{fig:ebtent_statistics_K14_M9}, but with $(K,M)=(14,10)$.}
        \label{fig:ebtent_statistics_K14_M10}
    \end{figure}
\end{centering}

One obvious solution is to add more descendants. If $D$ were to double to 32 (not shown), we would again recover a perfect tail estimate, not just in the aggregate but for \emph{for each family individually}. In statistical formalism, the MoCTail for the Tent Map with optimal AST is said to be a \emph{consistent estimator} of the truth, even with $N=1$: $\widehat{Q}(\nu,\mu;1,D,K-M)\to Q(\nu,\mu)$ as $D\to\infty$, for the simple reason that the van der Corput sequence from Eq.\ (\ref{eq:vandercorput}) approaches a uniform distribution over $[0,1)$, which propagates to a uniform distribution of descendant peaks over $(1-2^{-M},1)$, as shown by Eq.\ (\ref{eq:peak_bits_desc}), which is the climatological truth. The measurable discrepancy between the MoCTail and the truth actually vanishes at $D=2^{-M}/(\text{bin width})$ and integer multiples thereof. In general, as $D$ increases further, the estimator becomes less sensitive to $A$. Fig.\ \ref{fig:ebtent_statistics_K14_M10} shows results for $M=10$, in which case there are only $2^{13-10}=8$ bins to fill and the descendant pool does so easily---not just at $A=K-M=4$, but at $A=5$ when half of them have fallen out of the tail. 

The other solution besides increasing family size ($D$) is to increase the number of families ($N$) by extending the DS and then boosting the new ancestors as they appear. Whichever bins family 1 misses, family 2 might cover, and the tail coverage would improve as $N$ increases. We have empirical evidence (which will be presented later in section \ref{sec:eblogistic}), though not theoretical proof, that the MoCTail for the Tent Map is consistent for any fixed $D$ with increasing $N$.

The tradeoff between increasing $N$ and increasing $D$ is somewhat artificial for the Tent Map, but will become more fundamental and acute at the slightest extension to the Logistic Map below. To prepare for generalizing, we now reason through a principle for choosing AST that operates at a higher level than the bit expansion and may thus apply more broadly. 

The most obvious special feature of the Tent Map is its constant error growth rate, $|T'(z)|=2$ for all $z$. In contrast, the Logistic map's error growth rate $|L'(x)|$ ranges from 0 to 4, and more generally in nonlinear systems, predictability varies with the initial condition. Since each of the $N$ ancestors came from a different initial condition, each family might have a different optimal AST, $\{A_n\}_{n=1}^N$, which we call the ``conditionally optimal'' AST (\coast) as in \cite{Finkel2026boosting}. The pivotal question is: optimal in what sense? The KL objective is a \emph{joint function all $N$ ASTs}, so that in principle, all the $N$ optimization problems have to be coordinated. Worse, if we decide after an initial EB run that we need more ancestors, the optimal combination of ASTs might change. As a tractable, greedy, and parallelizable approximation, we insist on solving $N$ uncoupled optimization problems, treating each family as if it must approximate climatology on its own. That is, $\coast_n=\text{argmax}_A\tkl{\bp^{(n)}(A)}{\bq}$, where $\bp^{(n)}(A)$ is the descendant PMF arising from boosts of the $n$th ancestor from AST $A$ and $\bq$ is the true climatology. Explicitly, dropping the $n$ and $A$ dependence for brevity and introducing one more bin, $b=0$, to encompass the distribution body $(0,\mu]$: 
\begin{align}
    \label{eq:xckldef}
    p_b &= \frac1D\sum_{d=1}^D
    \begin{cases}
        \mathbb{I}\{z^*_d\leq\mu\} & \text{if }b=0\text{ (the fraction falling outside the tail)} \\
        \mathbb{I}\{\nu_{b-1}<z^*_d\leq\nu_b\} & \text{for }1\leq b\leq B \\ 
    \end{cases}\\
    \tkl{\bp}{\bq}
    &=
    \sum_{b=1}^B\bigg(\frac{p_b}{1-p_0}\bigg)\log\bigg(\frac{p_b/(1-p_0)}{q_b}\bigg).
\end{align}
The factor $1/(1-p_0)$ normalizes the tail of $\bp$. No such factor is needed for the truth $\bq$, because $\bq$ is a distribution over \emph{peaks above $\mu$}, with zero probability below $\mu$ by definition. Expanding the logarithm, we separate out the parts that depend on $\bq$:
\begin{align}
    \label{eq:elbo}
    \tkl{\bp}{\bq}&=\sum_{b=1}^B\bigg(\frac{p_b}{1-p_0}\bigg)\log\bigg(\frac{p_b}{1-p_0}\bigg)-\sum_{b=1}^B\bigg(\frac{p_b}{1-p_0}\bigg)\log q_b
\end{align}
We call the \emph{negative} of the first summation the ``Extreme-Conditional Entropy'' (\xclent), recognizing it as the Shannon entropy of the PMF of descendant peaks conditional on being in the tail. We posit the \xclent\ alone as a proxy objective function to maximize, since we do not know the terms $q_b$ in the second summation. The approximation has pros and cons, which section \ref{sec:discussion} will discuss, but presently we note that when all $q_b$s are constant---as they are for the Tent Map with evenly spaced bins ($q_b=1/B$)---the second term contributes a constant $+\log B$ and does not affect the optimization over $A$. We will call the AST maximizing \xclent\ the ``\xclent-\coast'', abbreviated $\astxc$ where convenient, to recognize that it only approximates the true \coast. 

Fig.\ \ref{fig:ebtent_storyline}d plots \xclent\ as a function of AST, alongside two other related entropies. Thresholded entropy, or \thrent, is the Shannon entropy of the unnormalized PMF $\{p_1,\hdots,p_B\}$, equivalent to \xclent\ with $p_0$ replaced by 0. It was introduced in \cite{Finkel2026boosting} to encourage both a high mean and a wide spread in descendant $z^*$s, as the maximum-entropy distribution is a uniform distribution. Lastly, total entropy or \totent\ is the Shannon entropy of the full PMF of $z^*$s, over $(0,1)$ with uniform bin spacing $2^{-13}$ throughout. We label the bins below $\mu$ as $\{p_{0,c}:1\leq c\leq C\}$ with $C=2^{13}\mu$ (their probabilities sum to $p_0$). The distribution's body, consisting of these non-tail bins, has its own entropy that we call \decent\ for ``Dull, Everyday Climatological Entropy''. All together, the entropies are defined
\begin{align}
    \thrent
    &= 
    -\sum_{b=1}^Bp_b\log p_b \\
    \decent
    &=
    -\sum_{c=1}^Cp_{0,c}\log p_{0,c} \\
    \totent 
    &=
    \thrent + \decent \\ 
    \xclent
    &= 
    \frac1{1-p_0}\big(\thrent\big)+\log(1-p_0) 
    & 
    \text{where } p_0
    =
    \sum_{c=1}^Cp_{0,c}
    =1-\sum_{b=1}^Bp_b
\end{align}
\xclent\ can also be interpreted as the extra entropy gained by resolving the tail into $B$ bins as opposed to lumping it all into one bin (normalized by the total tail weight):
\begin{align}
    \xclent
    &=
    \frac1{1-p_0}\Big(\totent - \big[\decent-(1-p_0)\log(1-p_0)\big]\Big)
\end{align}

In Fig.\ \ref{fig:ebtent_storyline}d, the three entropies \totent, \thrent, and \xclent\ all increase in lock-step, by one bit per increment of $A$ from 1 to 5. Both \thrent\ and \xclent\ peak at $A=K-M=5$, the same AST where the descendant scores uniformly fill the tail, and decline thereafter as descendants leave the tail. Meanwhile, \totent\ remains at the maximum value of 4 bits, seeing the descendants still occupying 16 separate bins counting those within the body. Fig.\ \ref{fig:ebtent_statistics_K14_M9}b shows that every ancestor's \xclent\ profile follows exactly the same pattern, peaking at $A=5$. The fact that KL divergence (Fig.\ \ref{fig:ebtent_statistics_K14_M9}d) reaches a minimum where \xclent\ reaches a maximum is validation of \xclent\ as a proxy objective. This fact is trivial for the Tent Map, but becomes actually useful when applied to the Logistic Map, where the theoretical prescription $A=K-M$ breaks down. 

\subsection{Ensemble Boosting on the Logistic Map}
\label{sec:eblogistic}

We now apply the same procedure as in section \ref{sec:ebtent} to the Logistic Map, but perturbing with the uniform distribution in $x$-space instead of $z$-space space (replacing all $z$s with $x$s in Eq.\ (\ref{eq:dzimplicit})). But crucially, we gerrymander bin boundaries to be evenly spaced in the $z$ variable (``$z$-regular'') rather than $x$-regular by mapping $\nu_0,\nu_1,\hdots,\nu_B$ through the conjugacy map $h(z)=\sin^2(\frac12\pi z)$, which means that $x$ bins get thinner towards the boundary. The threshold marking the start of the tail in $x$-space becomes $\mu=\sin^2\big(\frac12\pi(1-2^{-M})\big)$ to have the same exceedance probability as $z=1-2^{-M}$ in the Tent Map, which is $\mu\approx1-2^{-17}$ with $M=5$. The advantage of keeping $z$-regular bins is to make all $q_b$s constant, to satisfy the assumption that \xclent-\coast\ = \coast. It is certainly cheating by using knowledge of the truth $\bq$ to define the score $\kl{\bqhat}{\bq}$, but we press forward anyway with $z$-regular bins in order to change only one thing at a time: namely, perturbing $x$ instead of $z$, which snarles the orderly relationship we saw in the Tent Map between dynamical evolution and perturbation and makes the Logistic Map problem genuinely harder. Subsequently, in Section \ref{sec:discussion}, we will examine the impact of making the bins $x$-regular too.

\begin{centering}
    \begin{figure}
        \includegraphics[width=0.99\linewidth,trim={0cm 0cm 17cm 0cm},clip]{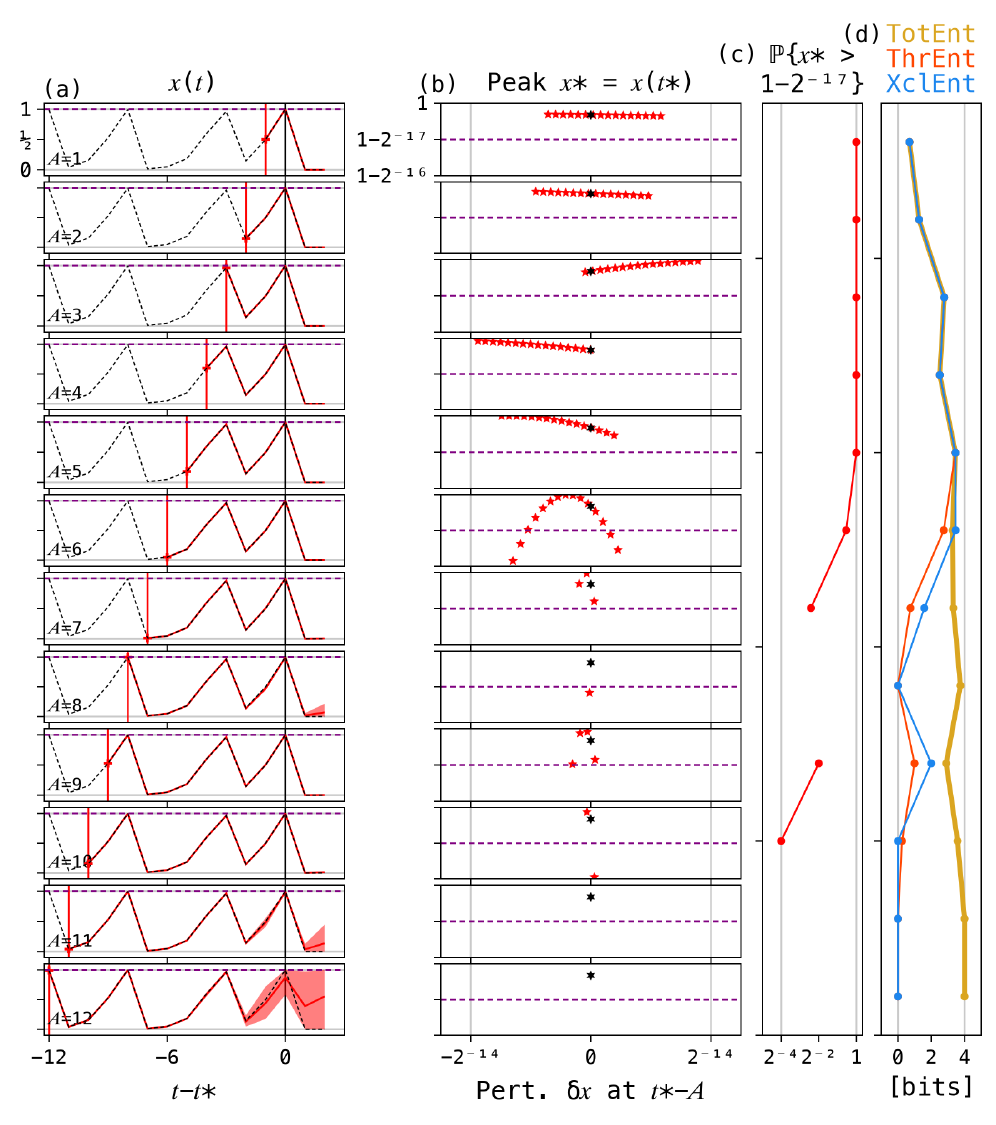}
        \caption{Ensemble Boosting of a single ancestor with the Logistic Map with $(K,M)=(14,9)$, formatted the same as Fig.\ \ref{fig:ebtent_storyline}.}
        \label{fig:eblogistic_storyline}
    \end{figure}
\end{centering}

\begin{centering}
    \begin{figure}
        \includegraphics[width=0.99\linewidth,trim={0cm 0cm 8cm 0cm},clip]{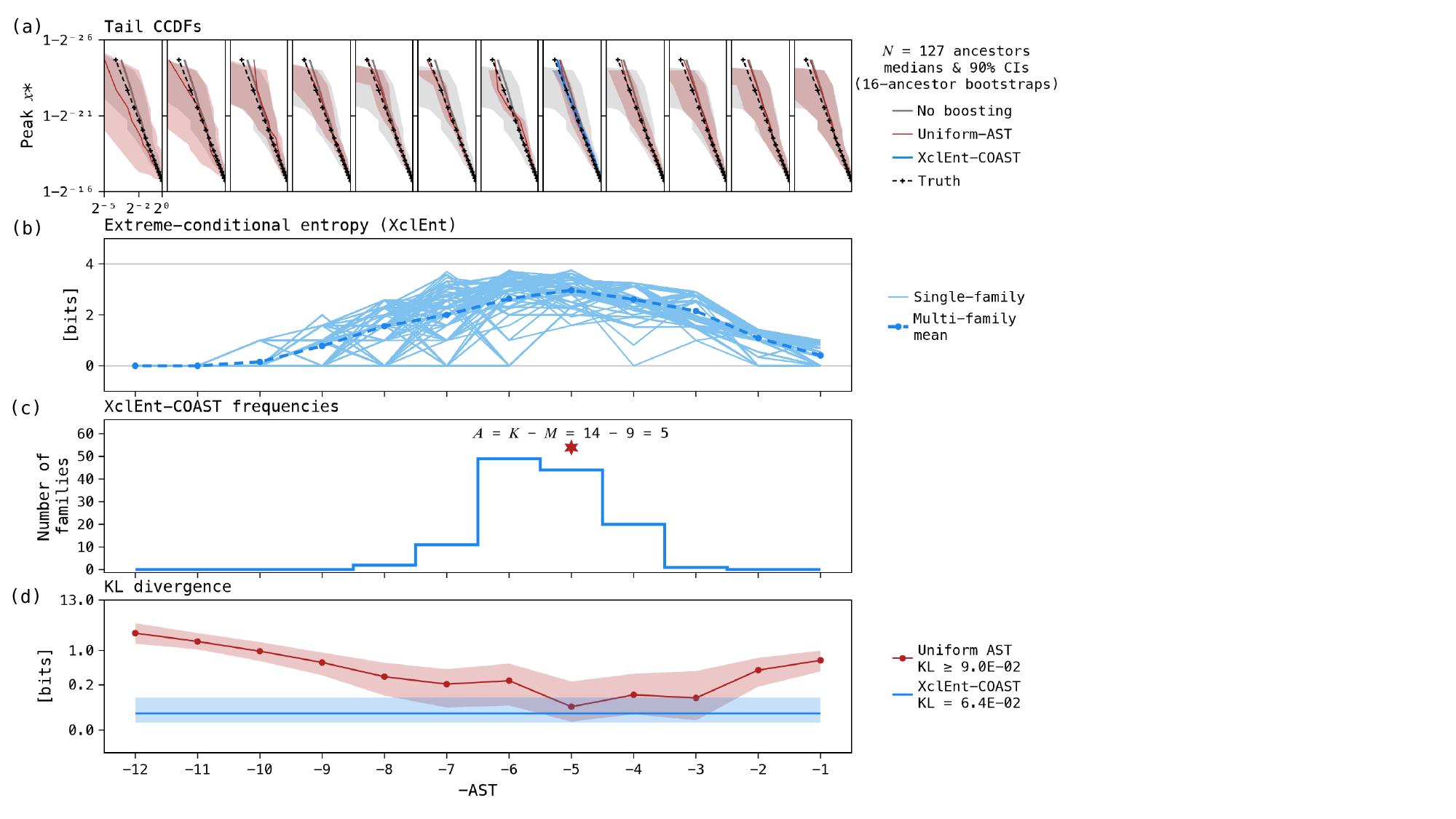}
        \caption{Performance of Ensemble Boosting with the Logistic Map with $(K,M)=(14,9)$, aggregated across ancestors, formatted the same as Fig.\ \ref{fig:ebtent_statistics_K14_M9} but with noticeable differences: (b) each family has a different \xclent\ as a function of $A$ , (c) there is a range of ASTs where they maximize (\xclent-\coast s), and (d) there is a nonzero spread in possible KL divergences when using \xclent-\coast\ as the AST selection rule.}
        \label{fig:eblogistic_statistics_K14_M9}
    \end{figure}
\end{centering}

\begin{centering}
    \begin{figure}
        \includegraphics[width=0.99\linewidth,trim={0cm 0cm 8cm 0cm},clip]{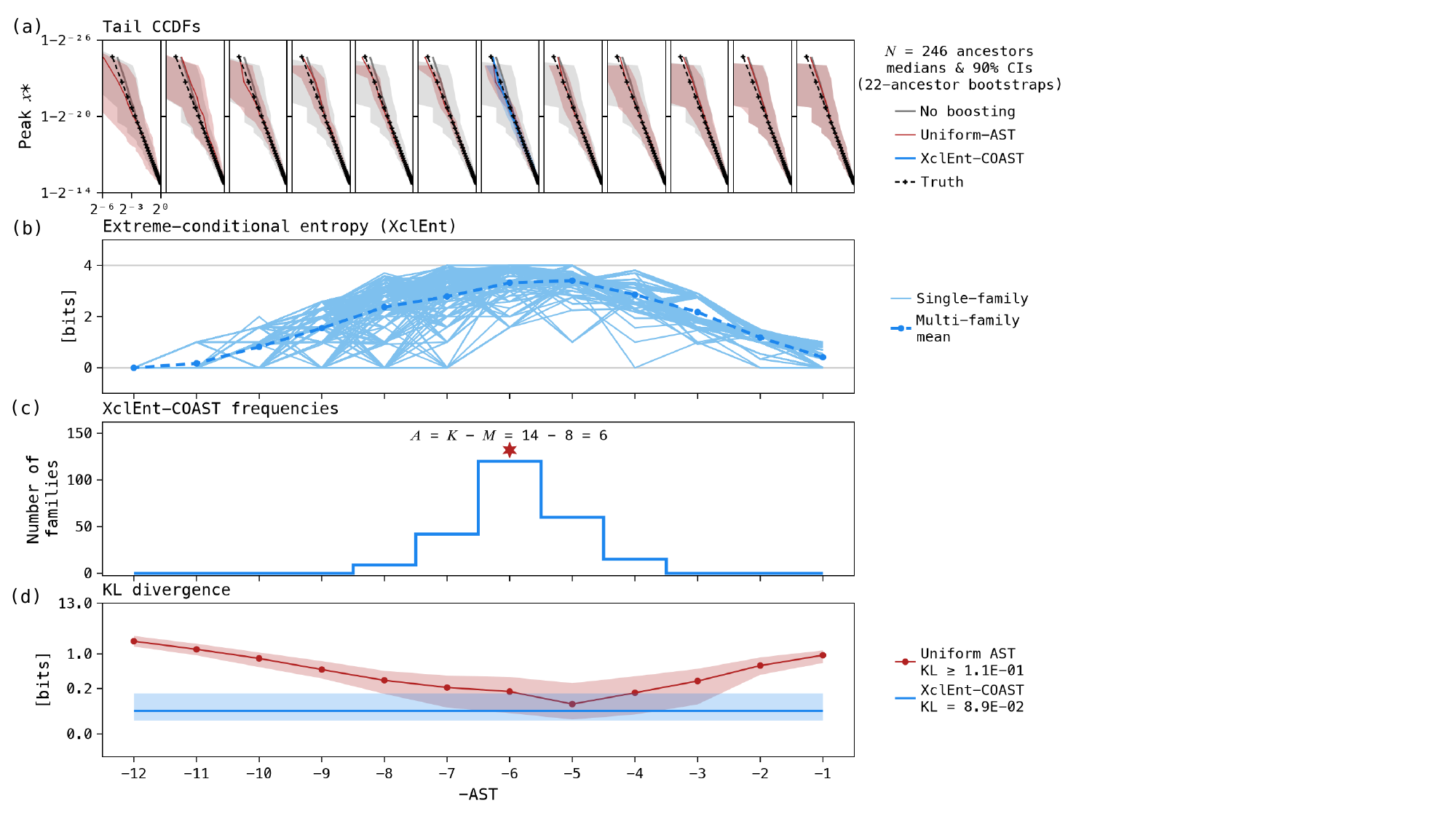}
        \caption{Same as Fig.\ \ref{fig:eblogistic_statistics_K14_M9}, but with $(K,M)=(14,8)$. }
        \label{fig:eblogistic_statistics_K14_M8}
    \end{figure}
\end{centering}

\begin{centering}
    \begin{figure}
        \includegraphics[width=0.99\linewidth,trim={0cm 0cm 8cm 0cm},clip]{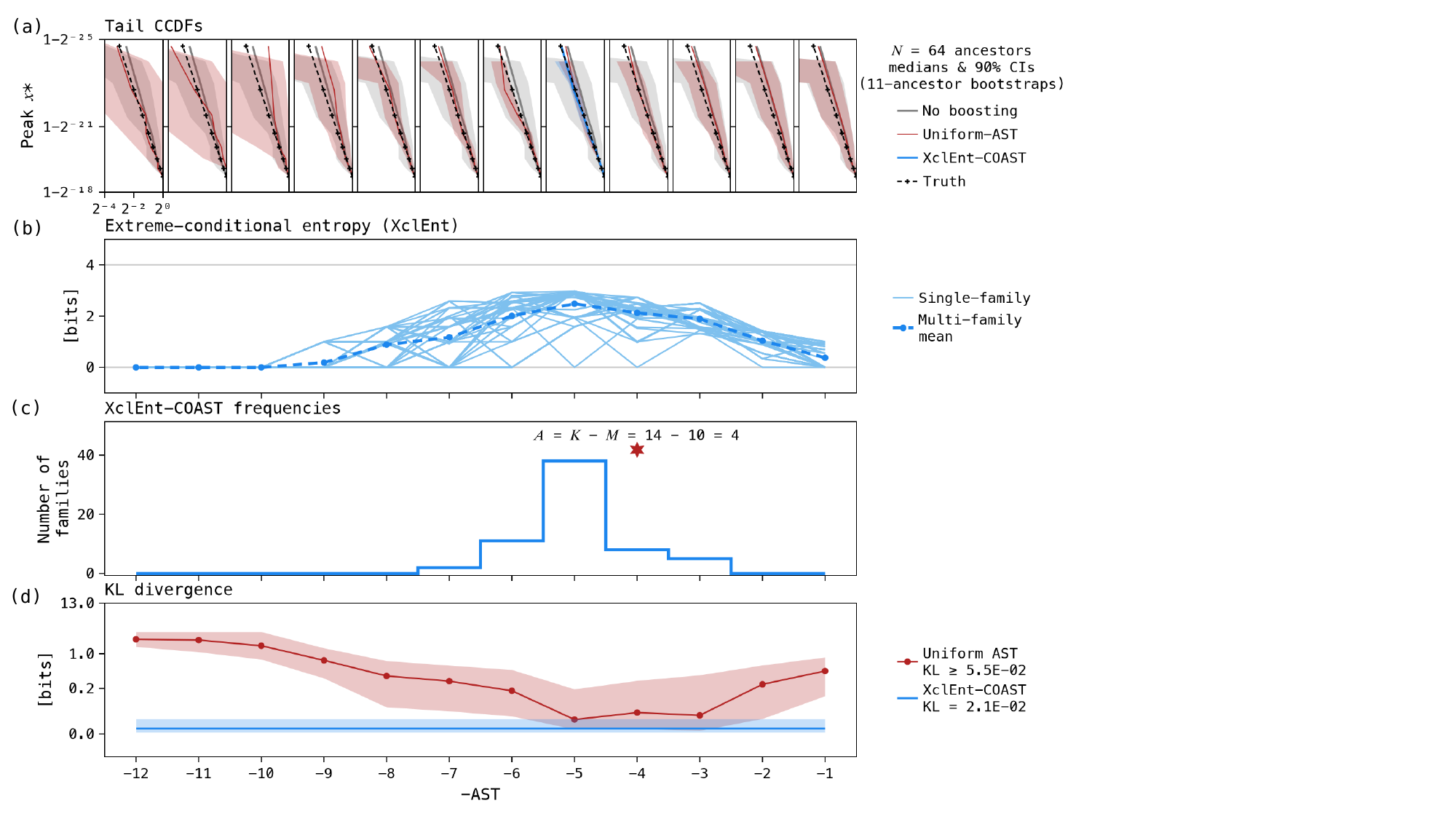}
        \caption{Same as Fig.\ \ref{fig:eblogistic_statistics_K14_M9}, but with $(K,M)=(14,10)$. }
        \label{fig:eblogistic_statistics_K14_M10}
    \end{figure}
\end{centering}
Fig.\ \ref{fig:eblogistic_storyline} shows the boosting procedure with the Logistic Map from a single ancestor. As in Fig.\ \ref{fig:ebtent_storyline}, the descendant scores (column b) spread out increasingly with longer $A$, but in a more abrupt way. As $A$ goes from 5 to 6, the function $x^*$ of $\delta x$ suddenly curves sharply downward, and descendant scores suddenly scatter widely, going from covering only the part of the tail to leaking substantially below it. It is difficult to judge by eye which AST is better, when faced with a tradeoff between diversity and population size inside the tail. No such tradeoff existed in the Logistic Map, in which descendants reach maximum diversity at $A=5$ and only afterward start exiting the tail, starting at $A=6$. The \xclent\ profile in Fig.\ \ref{fig:eblogistic_storyline} also indicates a tradeoff by never reaching its theoretical maximum value of $\log(D)=4$ bits, as it does for the Tent Map. Fig.\ \ref{fig:eblogistic_statistics_K14_M9}b shows that \xclent\ as a function of $A$ varies widely between ancestors (light blue lines), unlike the Tent Map case where they all looked identical. All of the profiles maximize somewhat below 4 bits of entropy, showing that incomplete tail coverage is typical. Moreover, the AST where they maximize (the \xclent-\coast) varies widely between ancestors, ranging from 3 to 8 according to the histogram in Fig.\ \ref{fig:eblogistic_statistics_K14_M9}c. The same patterns are apparent for other combinations of $(K,M)$; Figs.\ \ref{fig:eblogistic_statistics_K14_M8} and \ref{fig:eblogistic_statistics_K14_M10} show corresponding results for $(14,8)$ and $(14,10)$ respectively.

The fundamental reason that \xclent\ varies between ancestors is that $|L'(x)|$ and therefore the FTLE varies with $x$ (see Eq. \ref{eq:ftle_defn}), even though the asymptotic Lyapunov exponent is still $1$ 
\footnote{
    Using ergodicity, the Lyapunov exponent is the domain-average of $\log|L'(x)|$ under the stationary distribution $p_x$. Using $z=h(x)$, $T(z)=h\circ L(x)$, and $p_z(z)\,dz=p_x(x)\,dx$, it follows that $\mathbb{E}_{x\sim p_x}\log|\frac{dL(x)}{dx}|=\mathbb{E}_{z\sim p_z}\log|\frac{dT(z)}{dz}(\frac{dz}{dx}/\frac{dT(z)}{dL(x)})|=\mathbb{E}_z\log|\frac{dT(z)}{dz}|+\mathbb{E}_z\log|h'\circ h^{-1}\circ T(z)|-\mathbb{E}\log|h'\circ h^{-1}(z)|$.
    The last two expectations cancel out, since $h'\circ h^{-1}$ is just an example of an observable function whose expectation is preserved at stationarity, i.e., when $x\sim p_x$ (equivalent to $z\sim p_z$). So the \emph{global} Lyapunov exponent is preserved, but integration over the entire domain was crucial. 
    }.
Earth's atmosphere also has this feature, known as ``flow-dependent predictability'' \citep{Ferranti2015flow}, making it clear that AST generally should adjust for different initial conditions. 

How does our \xclent-\coast\ criterion fare at reconstructing the climatological tail? Fig.\ \ref{fig:eblogistic_statistics_K14_M9}d compares the performance, measured in KL divergence, of two AST selection rules: imposing the same \emph{unconditional} AST on all ancestors, denoted $\astuc$ (plotted in red as a function of $\astuc$), and customizing the AST as the \xclent-\coast, denoted $\astxc$ for each ancestor separately (plotted in blue as a constant, since \xclent-\coast\ is generated from a single rule and produces a single MoCTail). Among all choices for $\astuc$, $K-M=5$ is still the best choice in this example, but $\astxc$ is unambiguously better still. Its KL divergence from the truth is lower both in typical behavior and in variability. The tail CCDFs plotted in Fig.\ \ref{fig:eblogistic_statistics_K14_M9}a tell the same story, with the \xclent-\coast\ MoCTail plotted in blue following the truth better than any red line. (We overlay the blue in the panel corresponding to $\astuc=5$, which is the average \xclent-\coast\ across families according to the histogram in panel c, but stress that the \xclent-\coast\ mixture blends ASTs spanning across the range from 3 to 8). The corresponding Tent Map results (Fig.\ \ref{fig:ebtent_statistics_K14_M9}) are degenerate, as $\astxc=\astuc=K-M$ was found for all ancestors, leaving no difference between red and blue. State-dependent predictability is what makes it essential to condition AST on the ancestor.

But just how big is the advantage of $\astxc$ relative to $\astuc$, and either rule to DS, given an equal computational budget? The question is another version of the ``explore vs. exploit'' tradeoff that reinforcement learning grapples with: explore alternative routes into the tail by extending the DS to generate more ancestors, or exploit the initial conditions already discovered by riffing on the existing ancestors. Which strategy wins is hard to say \emph{a priori} in general, but some limiting cases are clear. If the computational budget affords ample ancestors ($N\sim 10\times$ the number of bins), then DS is the right choice because each sample is independent. If the budget is small, then boosting is probably a safer method to find new extremes, but only incrementally larger ones. Thus we compare their performances across a range of simulation costs, calculated as a function of the number of ancestors $N$:
\begin{align}
    \text{Cost of DS }&=N\times\big(\text{average return period}\big) \nonumber\\
    \text{Cost of EB with fixed AST $\astuc$}&=N\times\big(\text{average return period} + D\times\astuc\big)\\ 
    \label{eq:costformula}
    \text{Cost of EB with \xclent-\coast\ $\astxc$}&\approx N\times\big(\text{average return period} + D\times(K-M)\big) \nonumber.
\end{align}
Here, the mean return period is computed empirically from the DS, and is typically $\sim2^M$ but varies due to random sampling and the minimum imposed buffer time. We also use $K-M$ as a close approximation for mean \xclent-\coast, just for cost-counting purposes. 

\begin{centering}
    \begin{figure}
        \includegraphics[width=0.99\linewidth,trim={0cm 4cm 0cm 0cm},clip]{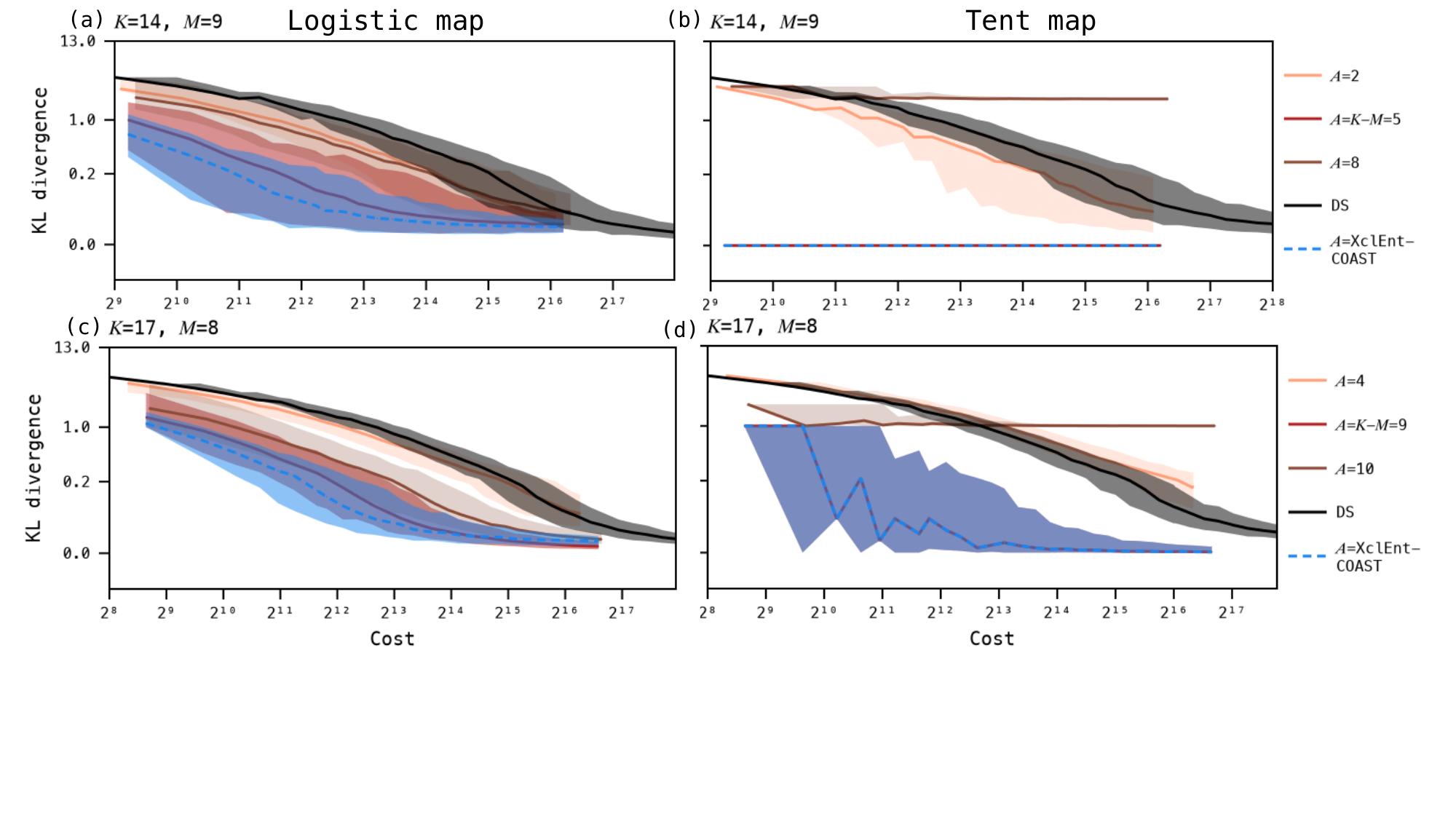}
        \caption{Performance vs. cost of different estimation methods. At a range of ancestor populations $N$, we calculate CCDFs using using boosting and the MoCTail estimator, using for the advance split time both the \xclent-\coast\ ($A=\astxc$, blue) and range of fixed ASTs ($A=\astuc$, orange for $\astuc<K-M$ and red for $\astuc=K-M$ and brown for $\astuc>K-M$). The center of the cost range, near $2^{13}$, is where the 16-ancestor results from Fig. \ref{fig:eblogistic_statistics_K14_M9} fall in panel (a). For comparison we also estimate the CCDF with DS, both with the ancestors without boosting (gray) and from an extended DS (black) for validation. Each method incurs some KL divergence, which is plotted as a function of its total cost. Total cost is linear in $N$, but with different constant factors for the different methods because of the overhead cost of boosting (see Eq. \ref{eq:costformula}). }
        \label{fig:klconv}
    \end{figure}
\end{centering}

Fig.\ \ref{fig:klconv} shows KL divergence as a function of cost for each method, revealing that different methods win in different situations. The small-budget regime (leftward on the plot) is where boosting wins most handily over DS---but only with an appropriate choice of AST. Short ASTs (yellow) perform similarly to DS, adding slightly more diversity to each ancestor at a slight extra cost. Very long ASTs (brown) are at best similar to the short ASTs, and at worst far inferior to DS, since perturbing initial conditions too far ahead of time erodes the peaks back to the climatological distribution and even shifts their timing, undermining any potential efficiency gains. Among all possible uniform-AST rules, $\astuc=K-M$ is always superior for the Tent Map (as we proved), but is also usually best for the Logistic Map. But allowing $A$ to be optimized per-ancestor unlocks greater gains: $A=\astxc$ generally outperforms all choices of uniform AST. The Tent Map has starker performance gaps than the Logistic Map, except for the perfectly optimal choice $\astxc\equiv\astuc=K-M$ which zeros out KL divergence when $M=9$ (no more bins than descendants). Performance gaps are narrower at lower thresholds and larger budgets. Even straightforward DS is competitive in that regime because new ancestors are relatively cheap to find. Another, perhaps equivalent perspective is that longer return periods imply more rarefied tails, which imply fewer routes into the tail according to Large Deviation Theory \citep{Galfi2021applications,Noyelle2023investigating}, which suggests that a small handful of ancestors should already capture all qualitatively distinct routes and that boosting can fill in the incremental variations. 

\begin{centering}
    \begin{figure}
        \includegraphics[width=0.99\linewidth,trim={0cm 4cm 0cm 0cm},clip]{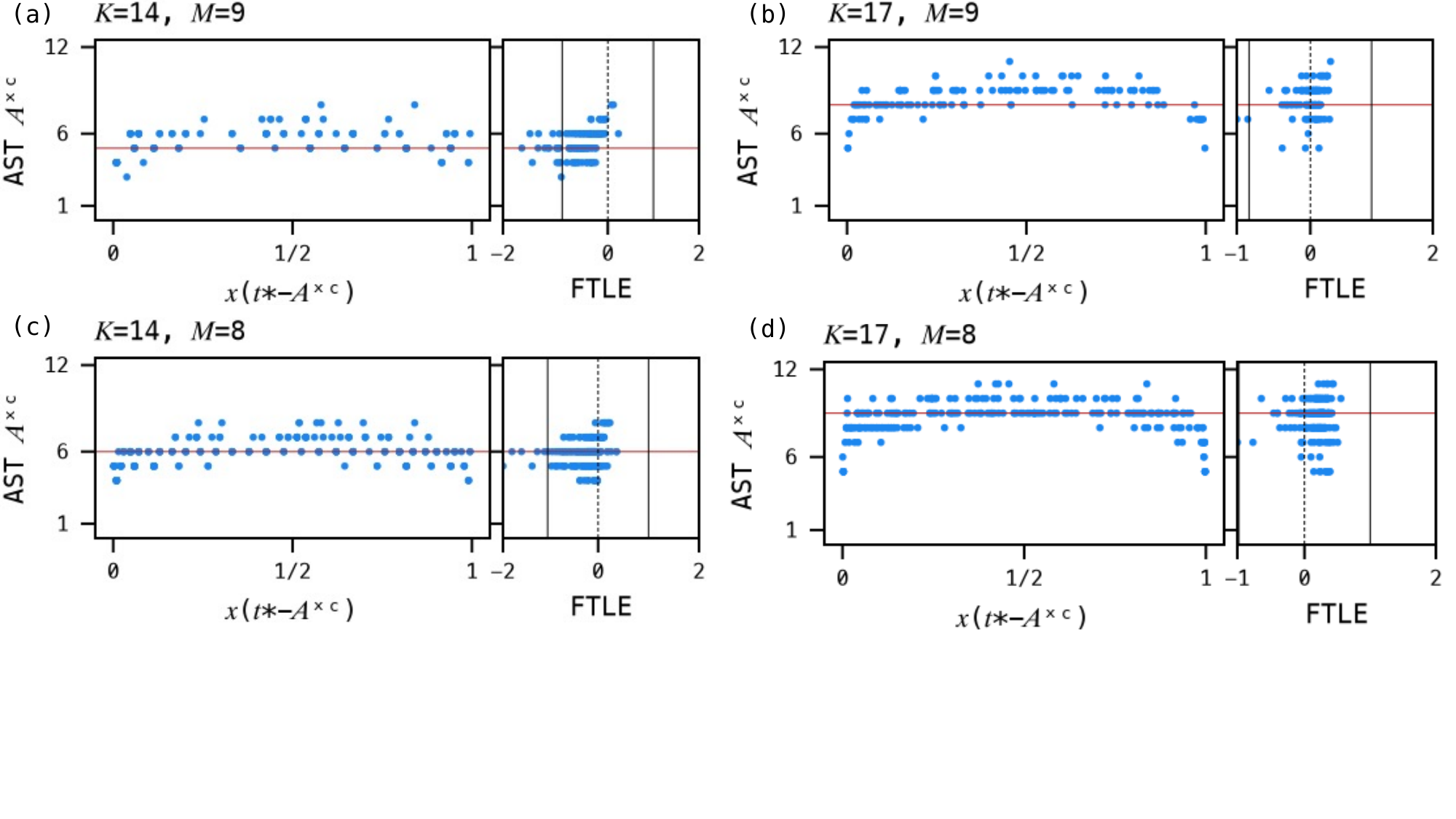}
        \caption{Initial condition dependence of \xclent-\coast\ in the Logistic Map, for four different combinations of $(K,M)$. For each peak $\{x(t_n^*)\}_{n=1}^N$ and corresponding value $\astxc$ of the \xclent-\coast, $\astxc$ is plotted as the ordinate, against the abscissa $x(t_n^*-\astxc)$ in the left panels, and the finite-time Lyapunov exponent (FTLE, expressed in base 2 as an inverse error-doubling time) in the right panels. The FTLE is calculated over the time horizon from $t_n^*-\astxc$ to $t_n^*$. }
        \label{fig:astofx}
    \end{figure}
\end{centering}

The cost estimates for $\astxc$ in Fig.\ \ref{fig:klconv} are optimistic because they disregard the cost of estimating the value of $\astxc$ for each ancestor. We envision mitigating this cost with adaptive optimization, either with adjoint-based methods \citep[e.g.,][]{Mu2003conditional,Vonich2024predictability} or adjoint-free methods such as Ensemble Kalman Inversion \citep{Kovachki2019ensemble} or the cross-entropy method \citep{deBoer2005tutorial}. Still, it is enticing to generalize Eq. (\ref{eq:coastformula}) accounting for state-dependent predictability on theoretical grounds. We search for some interpretable patterns in Fig.\ \ref{fig:astofx} by plotting $\astxc$ as a function of $x(t^*-\astxc)$ (the antecedent conditions identified as maximally \xclent-producing). There is an overall pattern of shorter \xclent-\coast\ near the domain boundaries, where $|L'(x)|$ is largest, which suggests that $\astxc$ might scale inversely with the FTLE. However, a scatterplot of $\astxc$ vs. FTLE shows no such decisive relationship. This is not surprising, as FTLE is a function of \emph{the entire trajectory segment} from $x(t^*-\astxc)$ to $x(t^*)$. A peak $x(t^*)$ is near 1 by definition, which means $x(t^*-1)$ is near $1/2$, where the error growth rate is near zero. $x(t^*-2)$ is then near either $0.146$ or $1-0.146=0.854$, where the growth rate is near 2.8,  and so on. The trajectory segment passes through regions of such widely varying error growth rates that the FTLE rapidly homogenizes. In light of this complexity, we don't pursue the theoretical extension of Eq. (\ref{eq:coastformula}) at present, but emphasize that the physical origins of $\astxc$ are a high-value target for rare event research.

\section{Discussion of arbitrary choices and generalizability}
\label{sec:discussion}

The Tent Map's special properties are now clearer in light of the preceding just-barely more complicated Logistic Map exercise, which gives a sense of the challenges for generalizing our EB recipe. 

\begin{enumerate}
    \item The Tent Map has no tradeoff between diversity (\thrent) and total population $(1-p_0)$ of descendants inside the tail: as AST increases, the descendant pool spreads out uniformly across the tail before any escape it. The Logistic Map, however, can in a single timestep scatter descendants from a small crowded region of the tail into a wide diaspora across the body, skipping over wide swaths of tail. Fundamentally, the reason relates to error growth: when a bit in the binary expansion flips somewhere beyond the $K$th position, the action of the Tent Map propagates the disturbance leftward by one position per time unit. The disturbance advances through the positions $[K+1,K,\hdots,A+M+1]$ which control \emph{where} the peak lands inside the tail, before having a chance to tamper with the positions $[A+M,A+M-1,\hdots,A+1]$ which controls \emph{whether} the peak lands inside the tail. On the other hand, the Logistic Map dynamics advances the disturbance erratically, by 0-2 bit positions with each time step due to variations in $|L'(x)|$ from 0.0-4.0. 
    \item The true climatological PDF is uniform for the Tent Map, which makes a ``good'' ensemble easy to recognize: it should be as uniform as possible across uniformly spaced bins in the tail. 
\end{enumerate}

To solve the Logistic Map exercise by only adjusting the AST (rather than transforming to the Tent Map, which effectively adjusts perturbation size and symmetry), we made two simplifying approximations. First, we chose to optimize the $A_n$s separately, tasking each family with the impossible goal of covering the tail single-handedly with its own PMF $\bp=\bp\ssn(A)$, which contributes to the MoCTail by $\bqhat=\frac1N\sum_{n=1}^N\bp\ssn(\astxc_n)$. Second, with $\tkl{\bp\ssn}{\bq}$ as the objective for the $n$th family, we relaxed the minimization problem for $\tkl{\bp\ssn(A)}{\bq}$ into the maximization problem for $\xclent_n(A)$, ignoring terms with $q_b$ in $\tkl{\bp\ssn(A)}{\bq}$ (second summation in Eq. \ref{eq:elbo}). The two problems are exactly equivalent if all $q_b$s are equal, which we ensured by gerrymandering bins to be $z$-regular. But this relied on knowing the very same tail we are trying to estimate! 

\begin{centering}
    \begin{figure}
        \includegraphics[width=0.99\linewidth,trim={0cm 0cm 8cm 0cm},clip]{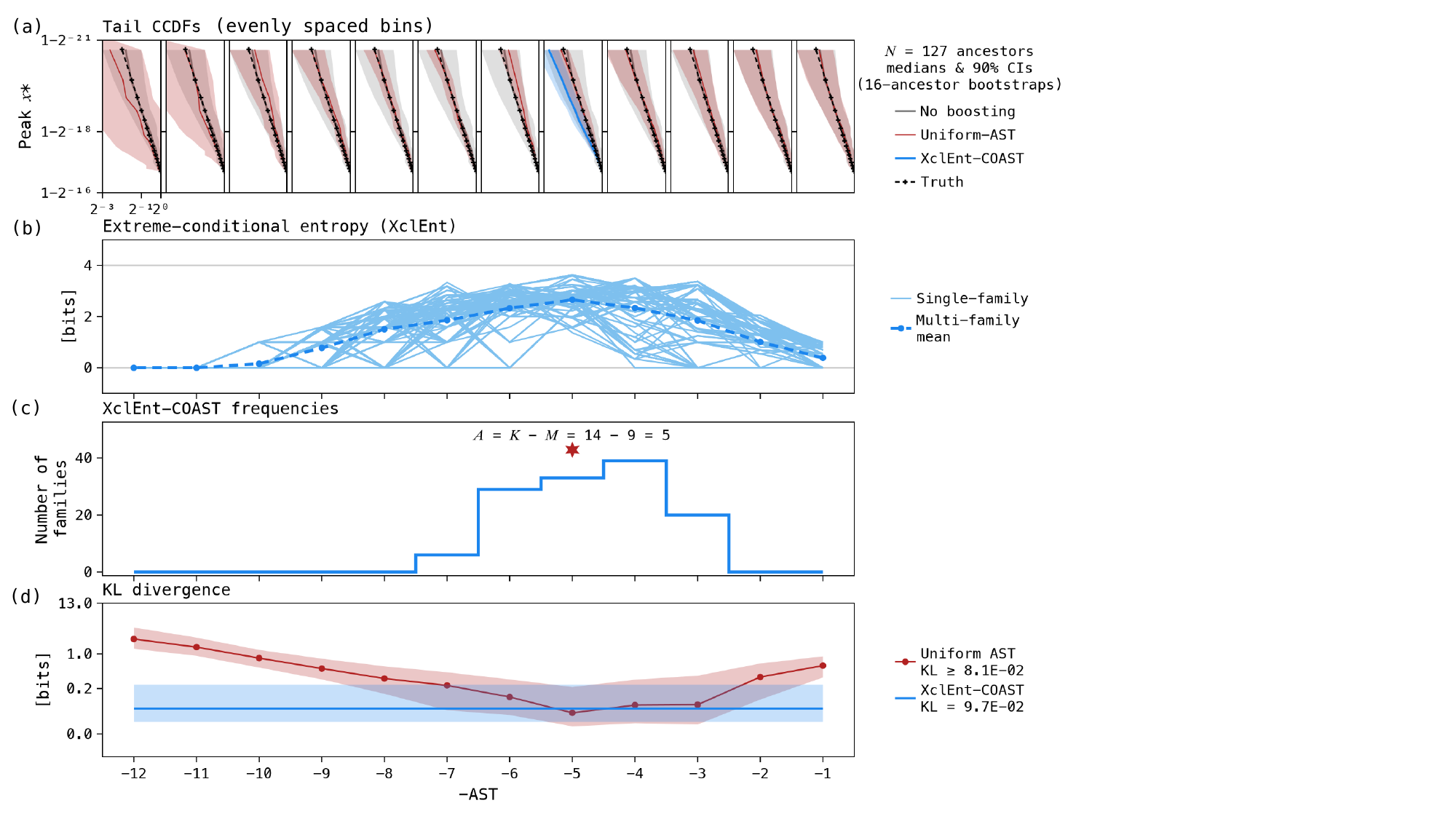}
        \caption{Performance of Ensemble Boosting with the Logistic Map with $(K,M)=(14,9)$, with bins uniformly spaced in $x$ ($x$-regular) rather than $z$-regular, where $x=\sin^2(\frac12\pi z)$. Figure format is the same as Fig.\ \ref{fig:eblogistic_statistics_K14_M9}.}
        \label{fig:eblogistic_statistics_K14_M9_unifbins}
    \end{figure}
\end{centering}

\begin{centering}
    \begin{figure}
        \includegraphics[width=0.99\linewidth,trim={0cm 0cm 0cm 0cm},clip]{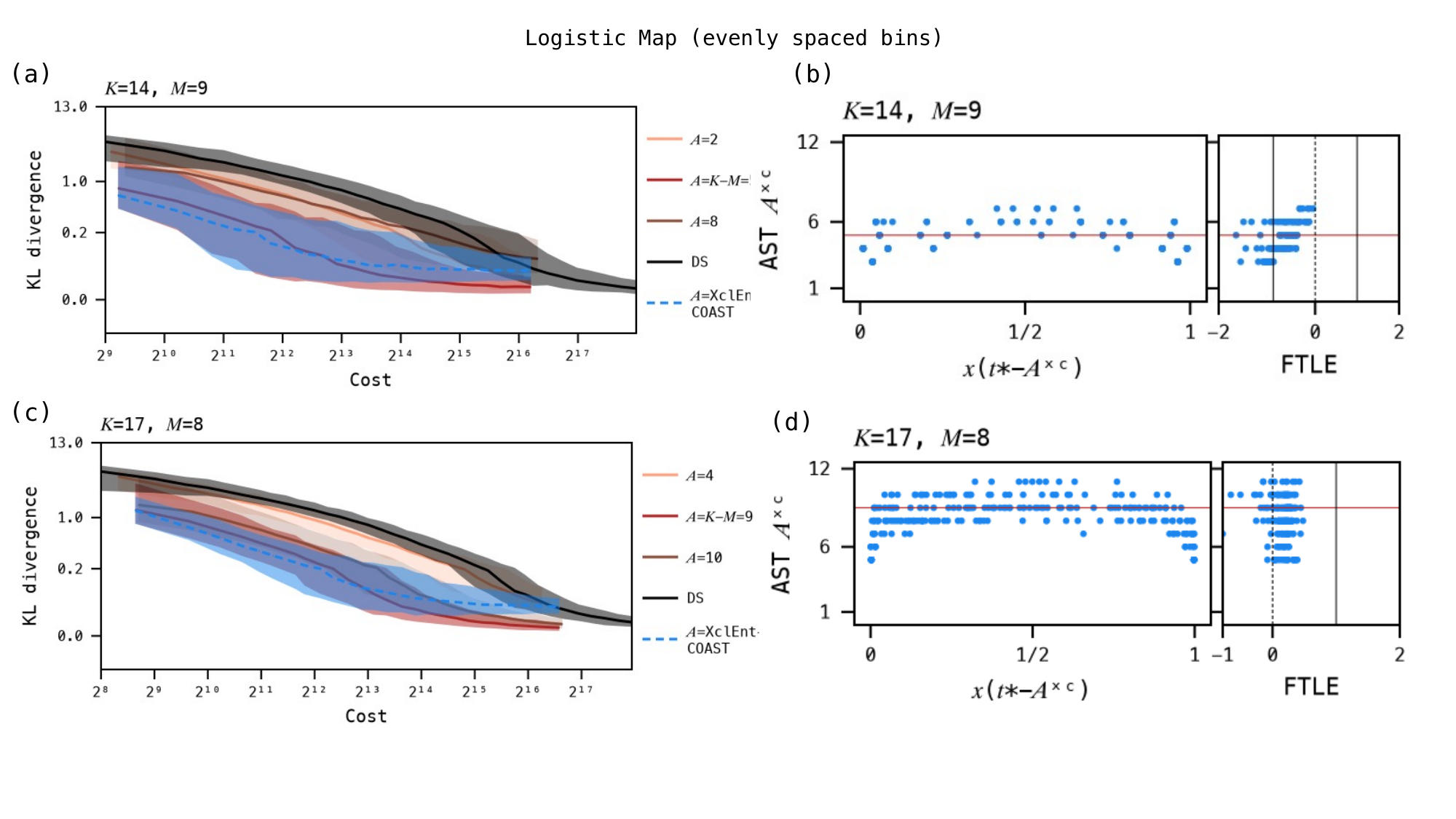}
        \caption{Results of Ensemble Boosting on the Logistic Map with $x$-regular bins, with $(K,M)=(14,9)$. (a,c) Performance vs. cost, formatted the same as Fig.\ \ref{fig:klconv}(a,c). (b,d) State dependence of the \xclent-\coast, formatted the same as Fig.\ \ref{fig:astofx}(a,c).}
        \label{fig:klconv_astofx_unifbins}
    \end{figure}
\end{centering}

Figs. \ref{fig:eblogistic_statistics_K14_M9_unifbins} and \ref{fig:klconv_astofx_unifbins} show that the plot thickens when bins are $x$-regular rather than $z$-regular. Error bands are wider and KL divergences are higher across the board, although numerical values aren't straightforwardly comparable with a different binning. The starker and more interesting difference is seen in Fig.\ \ref{fig:eblogistic_statistics_K14_M9_unifbins}a, where $\astuc=5$ now edges out $\astxc$ as the best AST rule. Their KL divergences are essentially tied (panel e), but the $\astxc$-based MoCTail seen in Fig.\ \ref{fig:eblogistic_statistics_K14_M9_unifbins} (blue in top row) systematically underestimates the truth. It makes sense when considering that \xclent\ favors uniformity across bins, whereas the true PMF puts more weight on the highest bins, thanks to the singularities of the Logistic Map's PDF at $x=0,1$ (Fig.\ \ref{fig:illustration_logisticmap}). It seems that choosing the same AST for all families somehow regularizes the MoCTail away from systematic underestimation. 

Yet we argue that \xclent\ is still a useful criterion for several reasons. Figs.\ \ref{fig:eblogistic_statistics_K14_M9}, \ref{fig:eblogistic_statistics_K14_M8}, and \ref{fig:eblogistic_statistics_K14_M10} as well as other $(K,M)$ combinations not shown suggest that both of the following averaged quantities are very accurate predictors of the optimal \emph{unconditional} AST $\astuc$:
\begin{itemize}
    \item The maximizer of mean \xclent s across ancestors, $\astuc=\text{argmax}_A\{\frac1N\sum_{n=1}^N\xclent_n(A)\}$, which is the $A$ at which the dark blue curve in Fig.\ \ref{fig:eblogistic_statistics_K14_M9_unifbins} (row 2) is maximized. 
    \item The mean of the $\astxc$s across ancestors, $\astuc=\overline{\astxc}=\frac1N\sum_{n=1}^N\text{argmax}_A\{\xclent_n(A)\}$, which is the mean of the histogram in Fig.\ \ref{fig:eblogistic_statistics_K14_M9_unifbins}c.
\end{itemize} 
In principle these can differ, but empirically they differ by at most 1. Fig.\ \ref{fig:klconv_astofx_unifbins} also shows that the \xclent\ criterion remains superior in the low-budget regime, up until some threshold cost near $2^{13}$ (Fig.\ \ref{fig:klconv_astofx_unifbins}a,c), but then plateaus in performance while the other methods continue to improve. In brief, \xclent\ appears efficient at small sample size but persistently biased at large sample sizes. 

The low-budget success of \xclent-\coast\ might be explained by a weaker, \emph{local} version of the uniform-$\bq$ approximation. Picture each pool of descendants sampling the tail as the fingers of a blind man grasping a small section of elephant. When the men are few and far between, like the sparse tail samples that a computer model produces within a truncated DS, a reasonable strategy to assemble the clearest possible elephant picture is for each man to explore as widely as possible, since the risk of bumping into each other or double-counting elephant sections is low. 

Mathematically, we capture the situation in three assumptions. 
\begin{enumerate}[label=(\roman*)]
    \item The $n$th family's boosting PMF $\bp\ssn=[p_b\ssn]_{b=1}^B$ is positive only on an ``accessible''  subset of bins $S\ssn\subseteq\{1,\hdots,B\}$, which cannot adjust with AST. This is a caricature of the scenario in Fig.\ \ref{fig:eblogistic_storyline}b, where the descendants spread out only within the top half of the tail while $A<5$ before hemorrhaging out of the tail at longer $A$.
    \item $q_b=q\ssn=$ constant for all $b\in S\ssn$, which is reasonable to expect if $S\ssn$ is a narrow contiguous block of bins and the true tail PDF is smooth. 
    \item Different families have non-overlapping boosting PMFs: $S\ssn\cap S\ssl=\varnothing$ if $n\neq\ell$. This is reasonable to expect when ancestors are very few in number, so likely sparsely distributed across the tail. 
\end{enumerate}
Then the full KL divergence can be decomposed over families ($n$) instead of bins ($b$):
\begin{align}
    \kl{\bqhat}{\bq}
    &=
    \sum_{b=1}^B
    \Bigg(\frac1N\sum_{n=1}^N\frac{p_b\ssn}{1-p_0\ssn}\Bigg)
    \log\Bigg(\frac1N\sum_{\ell=1}^N\frac{p_b\ssl}{1-p_0\ssl}\cdot\frac1{q_b}\Bigg) \\ 
    &=
    \frac1N\sum_{n=1}^N
    \sum_{b\in S\ssn}
    \Bigg(\frac{p_b\ssn}{1-p_0\ssn}\Bigg)
    \Bigg[\log\Bigg(\frac{p_b\ssn}{1-p_0\ssn} + \sum_{\ell\neq n}\frac{p_b\ssl}{1-p_0\ssl}\Bigg) - \log(Nq_b)\Bigg] \\
    &=
    \frac1N\sum_{n=1}^N\Bigg(-\xclent_n-\frac{\log(Nq\ssn)}{1-p_0\ssn}\sum_{b\in S\ssn}p_b\ssn\Bigg).
\end{align}
The last line drops terms in the ($\ell\neq n$) sum, which is zero by assumption (iii). Further, the second term in parentheses in the last line does not change with $A$ by assumptions (i) and (ii), so it can be dropped from the objective function. Ultimately, minimizing $\kl{\bqhat}{\bq}$ amounts to maximizing each $\xclent_n$ separately, recovering $\astxc$ as the proper target. 

Having now examined the validity and limitations of the various AST rules, we conclude this section by listing four practical strategies to leverage the best properties of each AST rule within a rare event sampling pipeline.

\begin{enumerate}
    \item Extend a DS piecemeal, boosting each peak over $\mu$ as it appears. Boost the early peaks at their individual $\astxc$s, which will deliver the best possible initial tail statistics. For later peaks, having seen enough examples to discern the distribution of $\astxc$s, switch to boosting at a uniform AST of $\astuc=\overline{\astxc}$, which will deliver the best long-term tail statistics. Transition smoothly between these strategies with some annealing schedule. 
    \item Actually attempt to construct $z$-regular bins by fitting a statistical model, such as a Generalized Extreme Value or Generalized Pareto distribution, to the early peaks \citep{Coles2001introduction}. Recall that the major added value of rare event sampling is not just to construct a return period curve, but to simulate realizations of the extreme events in their full spatiotemporal complexity. 
    \item Extend an EB pipeline piecemeal, refining bins sequentially in stages. For example, start by dividing the tail into two bins of equal estimated probability (which should be robustly estimable). After boosting to maximize \xclent\ over this two-bin PMF, the data should reveal where to place a new bin boundary to sub-divide whichever bin(s) are of interest, and then repeat boosting on the refined PMF, either by drawing fresh ancestors from the DS or boosting the first-stage descendants. Multi-stage boosting has been implemented before \citep{Gessner2021very}, and the same rough scheme resembles the TEAMS algorithm \citep{Finkel2024bringing,Finkel2026rare}. 
    \item Actually try to convert the problem into the Tent Map, or another system with the key desirable properties of uniform tails and uniform error growth that led to the exact prescription $A=K-M$. For example, one might train an autoencoder neural network to learn a conjugate mapping from the target dynamical variable, $x$, to a latent variable $z$, and thereby learn a state-dependent perturbation ($\delta x=\delta z|h'(x)|$) that precisely compensates for the state-dependent error growth. Machine learning of conjugate maps is already an active research field, recently demonstrated as a viable method for discovering unstable periodic orbits implementing chaos control algorithms \citep{Bramburger2021deep}. The sampling problem suggests a different set of loss terms, and we will pursue this idea in follow-up research. 
\end{enumerate}

\section{Conclusion}
\label{sec:conclusion}

To overcome the data-scarcity problem inherent to extreme events, the climate science community has taken up rare event sampling (RES) as a strategy to simulate more extremes more efficiently than straightforward model simulation can do. Strategically placed perturbations can trigger extreme events, and proper bookkeeping in the tilted ensemble can recover un-tilted statistics, enabling detailed risk analysis both qualitative and quantitative. RES is naturally framed as an ``applied'' toolset to enable risk analysis of hazardous weather with high societal impact, and understandably the trend has been to apply it to complex, coupled, expensive models. But there are fundamental open questions around the optimality of perturbations: what timing, structure, and magnitude of perturbation will deliver the most computationally efficient and statistically reliable sampling of the extreme events of interest? Dispatching these choices too hastily with heuristics can sacrifice major efficiency gains. 

In this paper, we have deliberately turned to the simplest possible one-dimensional chaotic systems to isolate and solve for one aspect of optimality: perturbation \emph{timing} within the framework of Ensemble Boosting (EB), a particular RES algorithm that is popular for its simplicity and flexibility. The closed formula that we derived and validated, though it applies strictly to only a specific toy system, relates directly to previously-used heuristics such as Lyapunov timescale and error saturation time and shows that neither is a complete answer without considering the level of severity being targeted. The new measure of ensemble spread that we derived, the extreme-conditional entropy (\xclent), provides a working proxy objective that can be used in Ensemble Boosting to choose conditionally optimal advance split times (\coast s). Although \xclent-\coast\ might be a biased estimator when applied independently across different initial conditions, it does seem to be optimally efficient in the regime of few initial conditions, which is a common practical setting. 

We have provided an exposition of a rare event algorithm and characterized its optimal parameters. The result is mathematically sound, and although it is planted within a toy sandbox, we have confidence that a more general theory can grow from it, in large part by drawing connections between ergodic theory, control theory, extreme value statistics, and atmospheric science. We eagerly invite active conversation and collaboration in the process.

\begin{acknowledgments}
J.F. gratefully acknowledges support from the University of Chicago through the Data Science Institute's AI for Climate (DSI-AICE) initiative, as well as the Institute for Climate and Sustainable Growth. Pedram Hassanzadeh, Dorian Abbot, and Paul O'Gorman provided invaluable advice and support. Many discussions with them as well as colleagues, especially Anna Asch, Alexander Wikner, and Zhixing Liu, helped to develop the ideas. 
\end{acknowledgments}

\section*{Data availability}
The \texttt{Julia} code used to run all the experiments and display results is publicly available in the Zenodo repository, COAST \citep[][\url{https://doi.org/10.5281/zenodo.20978271}]{justinfocus122026COAST}. Interested readers are encouraged to contact J.F. for using the code.

\bibliography{references_cap}

\end{document}